\documentclass[iop]{emulateapj}
\usepackage{natbib}
\usepackage{longtable}
\usepackage{rotating}

\begin{document}
\title{Hot Subdwarf all southern sky fast transit survey\\ with the Evryscope}

\author{Jeffrey~K.~Ratzloff\altaffilmark{1}, Brad~N.~Barlow\altaffilmark{2}, P\'{e}ter~N\'{e}meth\altaffilmark{3,4}, Henry~T.~Corbett\altaffilmark{1}, Stephen~Walser\altaffilmark{2}, Nathan~W.~Galliher\altaffilmark{1}, Amy~Glazier\altaffilmark{1}, Ward~S.~Howard\altaffilmark{1}, Nicholas~M.~Law\altaffilmark{1}}

\altaffiltext{1}{Department of Physics and Astronomy, University of North Carolina at Chapel Hill, Chapel Hill, NC 27599-3255, USA}
\altaffiltext{2}{Department of Physics and Astronomy, High Point University, High Point, NC 27268, USA}
\altaffiltext{3}{Astronomical Institute of the Czech Academy of Sciences, Fri\v{c}ova 298, CZ-251\,65 Ond\v{r}ejov, Czech Republic}
\altaffiltext{4}{Astroserver.org, 8533 Malomsok, Hungary}

\email[$\star$~E-mail:~]{jeff215@live.unc.edu}


\begin{abstract}

We have conducted a survey of candidate hot subdwarf stars in the southern sky searching for fast transits, eclipses, and sinusoidal like variability in the Evryscope light curves. The survey aims to detect transit signals from Neptune size planets to gas-giants, and eclipses from M-dwarfs and brown dwarfs. The other variability signals are primarily expected to be from compact binaries and reflection effect binaries. Due to the small size of hot subdwarfs ($R \approx 0.2 R_{\odot}$), transit and eclipse signals are expected to last only $\approx$twenty minutes, but with large signal depths (up to completely eclipsing if the orientation is edge on). With its 2-minute cadence and continuous observing Evryscope is well placed to recover these fast transits and eclipses. The very large field of view (8150 sq. deg.) is critical to obtain enough hot subdwarf targets, despite their rarity. We identified $\approx$11,000 potential hot subdwarfs from the 9.3M Evryscope light curves for sources brighter than $m_{g}$ = 15. With our machine learning spectral classifier, we flagged high-confidence targets and estimate the total hot subdwarfs in the survey to be $\approx$1400. The light curve search detected three planet transit candidates, shown to have stellar companions from follow--up analysis. We discovered several new compact binaries (including two with unseen degenerate companions, and several others with potentially rare secondaries), two eclipsing binaries with M-dwarf companions, as well as new reflection effect binaries and others with sinusoidal like variability. The hot subdwarf discoveries identified here are spectroscopically confirmed and we verified the Evryscope discovery light curve with TESS light curves when available. Four of the discoveries are in the process of being published in separate followup papers, and we discuss the followup potential of several of the other discoveries.

\end{abstract}


\section{INTRODUCTION} \label{section_intro}

Hot subdwarfs (HSD) are small, dense, under-luminous, high temperature stars. Most are thought to be helium cores with a thin hydrogen layer, formed from stripping of the main hydrogen shell during the red-giant phase by a binary companion. The hydrogen stripping is believed to prevent the core collapse, outer layer ejection, and degenerate remnant associated with the typical post red giant cycle. Instead the HSD will be a stable, helium core burning star that is underluminous for its temperature. A thorough analysis of the formation of HSDs via binary interaction can be found in \cite{2002MNRAS.336..449H, 2003MNRAS.341..669H}. HSDs are observed with temperatures typically in the 25,000 K to 40,000 K range and with a small radius and mass ($R \approx 0.2 R_{\odot}$ and $M \approx 0.5 M_{\odot}$). A comprehensive review of HSDs can be found in \cite{Heber_2016}.

Given this evolutionary theory, most HSDs are thought to have companions, with observations generally supporting this idea \citep{2001MNRAS.326.1391M, 2004Ap&SS.291..321N, 2011MNRAS.415.1381C}, although there is a non-trivial fraction ($\approx 1/3$) of observed single HSDs that are challenging to explain. HSD are observed with companions ranging from white dwarfs up to F stars, and periods from a few hours to several years. HSD binaries include compact degenerate systems, with a few massive systems thought to be potential supernovae progenitors \citep{2000MNRAS.317L..41M, 2007ASPC..372..393G, 2012ApJ...759L..25V, 2013A&A...554A..54G}, and a handful of peculiar systems thought to be very rare merger candidates \citep{2017ApJ...835..131K, 2019ApJ...883...51R}. Compact HSD systems can also be found with late stellar or brown dwarf companions. The eclipsing type are designated as HW Virs (for a complete list of known solved systems see \citealt{2018A&A...614A..77S}), and two examples of non-eclipsing, reflection effect systems can be found in \cite{2014A&A...570A..70S}. Wider systems with K-type, G-type, and earlier main sequence companions have also been discovered; a proven approach uses photometric color data \citep{2003AJ....126.1455S, 2004PASP..116..506R} to identify likely composite sources. Spectroscopic work by \cite{2018MNRAS.473..693V} revealed some long-period systems with demonstrated double line spectra. Planet companions are thought possible, with a few circumbinary candidates, although none have been demonstrated conclusively. An interesting chronicle of HSD circumbinary planet hunting can be found in \cite{Heber_2016}. The rich extent of HSD variability allows for testing of formation and evolution theory, and for careful measurement of HSD properties.

HSD binaries are generally placed into two groups based on the nature of the companion interaction during the formation process. Progenitor systems with comparatively smaller and closer companions are thought to be unable to accrete matter (from the hydrogen shell of the red-giant, HSD progenitor) at a fast enough rate to be stable. Referred to as a common envelope (CE), the CE phase will result in matter being ejected during the mass transfer with a resulting loss in angular momentum of the system and a tightening of the binary period. A description of the HSD formation CE channel can be found in \cite{2008MmSAI..79..375H}. Post CE HSD binaries typically have periods from 2 hours up to 30 days, with a few known exceptionally short period systems. Common companions are M-dwarfs, K-dwarfs, and white dwarfs; although more exotic remnant companions are possible. Progenitor systems with larger and farther companions form the second group of HSD binaries as they are thought to be able to accrete matter at a sufficient rate to avoid substantial mass ejection. This Roche Lobe Overflow (RLOF) formation is credited with producing wider HSD systems \citep{2013MNRAS.434..186C, 2019MNRAS.482.4592V}, containing earlier (G and earlier) main sequence companions with typical periods between 400 and 1600 days. There is a period "gap" between 30 and 300 days with few observed systems in this period range, likely due to differences in the two formation channels. 

We have conducted an all-southern-sky (all RA, Dec $< +10^\circ$), bright ($m_{V} < 15$) HSD survey aimed at finding post CE phase binaries and variables, as well as transiting planets. We use the fast, 2-minute cadence photometric observations from the Evryscope to look for periodic signals in the light curves. The wide-seeing (8150 sq. deg. instantaneous field-of-view (FoV)) Evryscope is a gigapixel-scale telescope that is optimized to find rare, fast transit objects (including compact binaries, short period eclipsing binaries, and planet transits lasting only tens of minutes or less). It is designed for short-cadence observations with continuous all sky coverage and a multi-year-period observation strategy \citep{2015PASP..127..234L, 2019PASP..131g5001R}. Most importantly for the HSD search, the Evryscope is highly sensitive to the observationally challenging, approximately twenty-minute duration transits and eclipses expected from HSDs. The continuous, 2-minute Evryscope images ensure the transits are well sampled even at the shortest expected periods. The wide FoV and continuous observing provides light curves for enough bright sources (9.3M with $m_{g} < 15$M), that we have a substantial number of HSD targets for our survey (several thousand), despite their rarity. The multi-year observing strategy provides tens of thousands of epochs per target, increasing the chance of capturing enough fast transits to enable detections. Our survey covers periods from 2-720 hours, with typical sensitivity to few-percent level variation.

As a complement to the Evryscope light curves, we developed a machine-learning based spectral classifier to help identify potential HSD targets in the Evryscope database, and to provide a confidence level to prioritize discovery followup. A subset of targets is spectroscopically confirmed as a test of the HSD target list performance, and to more accurately estimate the total HSD targets in the survey. The homogeneous, single instrument light curve dataset helps greatly in our estimation of the survey sensitivity, which we combine with the classifier results to estimate occurrence rates for several of the HSD binary types.

The HSD survey in this work identified 117 variables with 79 known and 38 new discoveries (including 14 HSDs). Two of the new discoveries are compact binaries showing strong light curve variation due to ellipsoidal deformation effects from an unseen degenerate companion. Two others are bright, new HW Vir discoveries. The peculiar variability of these systems was a key factor in their discovery, and demonstrates an advantage of the light curve driven HSD survey approach. We also detected 3 planet transit candidates, later shown to be stellar companions. We found several reflection effect HSD binaries, and others with sinusoidal like variability. The survey revealed several other potentially high-priority targets for followup, which we discuss in \S~\ref{section_discovery}. See Table \ref{tab:discoveries_sum_tab} for a summary of the discoveries in this work.

\begin{table}[h]
\caption{Survey Detections}
\begin{tabular}{ l c c c}
Detection & HSD$^a$ & Other$^b$ & Total\\
\hline
New Discoveries & 14 & 24 & 38\\
Known Recoveries & 14 & 65 & 79\\
\hline
Total & 28 & 89 & 117\\
\hline
\multicolumn{4}{l}{$^a$Spectroscopically confirmed HSDs.}\\
\multicolumn{4}{l}{$^b$Other stellar type than HSD.}\\
\end{tabular}
\label{tab:discoveries_sum_tab}
\end{table}

This paper is organized as follows. In \S~\ref{section_obs} we describe the observations leading to the discoveries as well as the variability search including the generation of the target list, survey coverage and estimated number of HSDs. In \S~\ref{section_det_of_var} we detail the detection process and expected recovery based on transit simulations. We show the followup observations in \S~\ref{section_followup} including identification spectra, radial velocity for a select target, and confirmation light curves. The discoveries from the survey are shown in \S~\ref{section_discovery}, along with the best fit to the photometric variability and ID spectra. We also discuss unique features and characteristics of the discoveries, and suggest additional followup. We discuss the survey sensitivity and the potential for a followup survey in \S~\ref{section_discussion}, and conclude in \S~\ref{section_summary}.


\section{OBSERVATIONS AND VARIABILITY SEARCH} \label{section_obs}

\subsection{Evryscope Photometry}

The hot subdwarf survey in this work is based on Evryscope photometric observations taken from January, 2016 to June, 2018. The exposure time was 120 s through a Sloan {\em g} filter providing an average of 32,600 epochs per target. The wide-seeing Evryscope is optimized to find rare, fast transit objects. It is a robotic 22 camera (each with 29MPix) array mounted into a 6 ft-diameter hemisphere which tracks the sky \citep{2015PASP..127..234L, 2019PASP..131g5001R}. The instrument is located at CTIO in Chile and observes continuously, covering 8150 sq. deg. in each 120s exposure. The Evryscope monitors the entire accessible Southern sky at 2-minute cadence, providing tens of thousands of epochs on 16 million sources (with 9.3M sources brighter than 15M in $m_{g}$).

Here we only briefly describe the calibration, reduction, and extraction of light curves from the Evryscope; a detailed description can be found in the Evryscope instrumentation paper \citep{2019PASP..131g5001R}. Raw images are filtered with a quality check, calibrated with master flats and master darks, and have large-scale backgrounds removed using the custom Evryscope pipeline. Forced photometry is performed using APASS-DR9 \citep{2015AAS...22533616H} as our master reference catalog. Aperture photometry is performed on all sources using multiple aperture sizes; the final aperture for each source is chosen to minimize light curve scatter. Systematics removal is performed with a custom implementation of the SysRem \citep{2005MNRAS.356.1466T} algorithm.

\subsection{Evryscope Target Search Lists} \label{section_classifier_list}

\subsubsection{Hot Subdwarfs as a Spectral Type}

The HSDs in this work are defined as a spectral type, with the initial selection chosen by color / magnitude space and the final determination decided by surface gravity and temperature obtained from followup spectra. This approach includes the traditional sdB, sdO, and other variants (sdOB, He-sdB, He-sdO), all understood to be evolutionary track driven HSDs. Also included are some extreme horizontal branch (EHB), blue horizontal branch (BHB), post asymptotic giant branch (AGB), and transitioning objects passing through the color magnitude space occupied by sdB and sdO HSDs. The surface gravity and temperature requirements for our sdB and sdO discoveries are $\log g > 4.8$ and $T_{\rm eff} > 20,000 K$, with other exotic HSD discoveries (pre-He WD, BHB, or post-AGB) designated as distinct objects.

\subsubsection{The Evryscope Hot Subdwarf Search List}

The Evryscope hot subdwarf target search list is a combination of four sources: two published lists and two internally generated lists to match our light curve database. All lists are generated using a form of color / color or color / magnitude parameter space selection. Each approach has differences in the data or selection method used, that provide a confidence level (recovery and false positive estimates) unique to the particular approach. Here we define the confidence levels, which we demonstrate in the following sections and use to estimate the number of HSD targets in our survey, as well as to prioritize HSD variable candidates for further followup.\\

\textbf{Very High Confidence Level:} Target is a member of 3 or 4 of the search lists.\\
\textbf{High Confidence Level:} Target is a member of both of the GAIA-DR2 \citep{2018A&A...616A...1G} based search lists.\\
\textbf{Medium Confidence Level:} Target is a member of one of the GAIA-DR2 based search lists.\\
\textbf{Global Confidence Level:} All Targets in the survey regardless of origin.\\

See the following sections for further search list generation details.

\subsubsection{Machine-Learning Generated Search Lists} \label{section_gen_lists}

The two internally generated lists for the hot subdwarfs are:\\
1) A machine-learning based stellar classifier (hearafter the Evryscope Classifier) we developed based on publicly available data from APASS \citep{2015AAS...22533616H} and PPMXL \citep{2010AJ....139.2440R}, which we use to select hot subdwarf candidates. The Evryscope Classifier is a multi-step machine-learning algorithm that uses reduced proper motion and B-V color differences to determine stellar size and spectral type. When available, we use additional color differences (V-K, J-H, H-K) to determine the luminosity class.\\
2) A modified version of the Evryscope Classifier that uses GAIA-DR2 \citep{2018A&A...616A...1G} data (Evryscope GAIA Classifier), with a similar machine-learning approach but with absolute G magnitude (parallax corrected) and B-R color differences.\\
Filtering the lists to match our field-of-view (Dec $< +10$) and magnitude range ($m_{V} < 16$), provides 10,892 and 5957 targets respectively.

\subsubsection{Published Search Lists} \label{section_published_lists}

The two external lists for the Hot Subdwarfs are:\\
1) \cite{2017OAst...26..164G}, a composite source based Hot Subdwarf candidate list.\\
2) A GAIA-DR2 \citep{2018A&A...616A...1G} based Hot Subdwarf candidate list \citep{2019A&A...621A..38G}.\\
We filter the lists to match our field-of-view (Dec $< +10$) and magnitude range ($m_{V} < 16$), yielding 1900 and 5963 targets respectively.

\subsubsection{Evryscope Classifier} \label{section_es_class}

We developed a machine-learning based classifier that uses publicly available catalog data to estimate stellar size from a B-V color/magnitude space, and to estimate spectral type from multiple color-differences. All sources in Evryscope database were matched to APASS-DR9 \citep{2015AAS...22533616H} and PPMXL \citep{2010AJ....139.2440R} catalogs to obtain reduced proper motion (RPM) and color differences (B-V, V-K, J-H, H-K) for each target. With:

\begin{equation}
    RPM = M_{V} + 5 \log(\sqrt{(pm_{ra}^2+pm_{dec}^2)/1000}) + 5
\end{equation}

Modifying the method in \cite{1972ApJ...173...xxx} with a two step machine learning process described below, we classify stars based on B-V and RPM to identify stellar size - main sequence, giants, white dwarfs, or subdwarfs.The RPM and B-V combination provides a high return on our target catalog ($\approx$ 99\% of our targets are classified) and captures spectral information using available data. After the stellar size estimation is completed, the four color differences are used to approximate the spectral type.

\begin{figure}[h!]
\centering
\includegraphics[width=1.0\columnwidth]{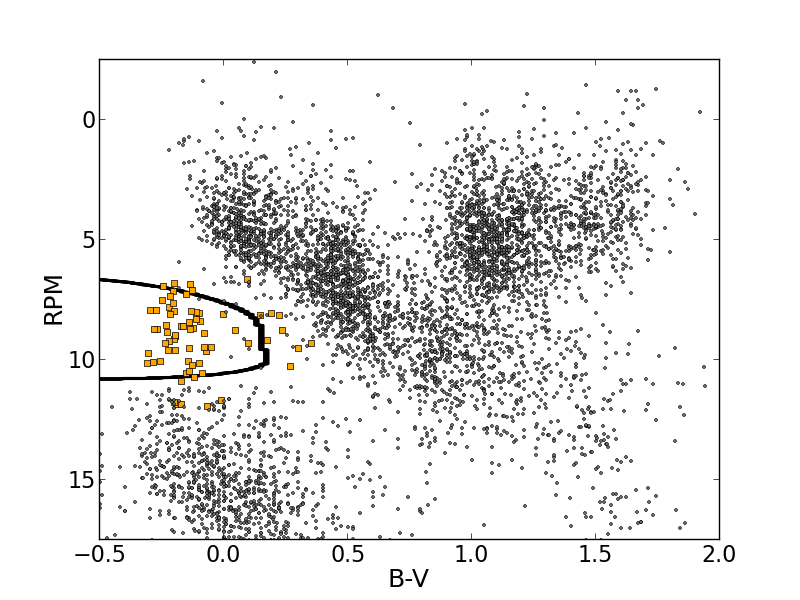}
\includegraphics[width=1.0\columnwidth]{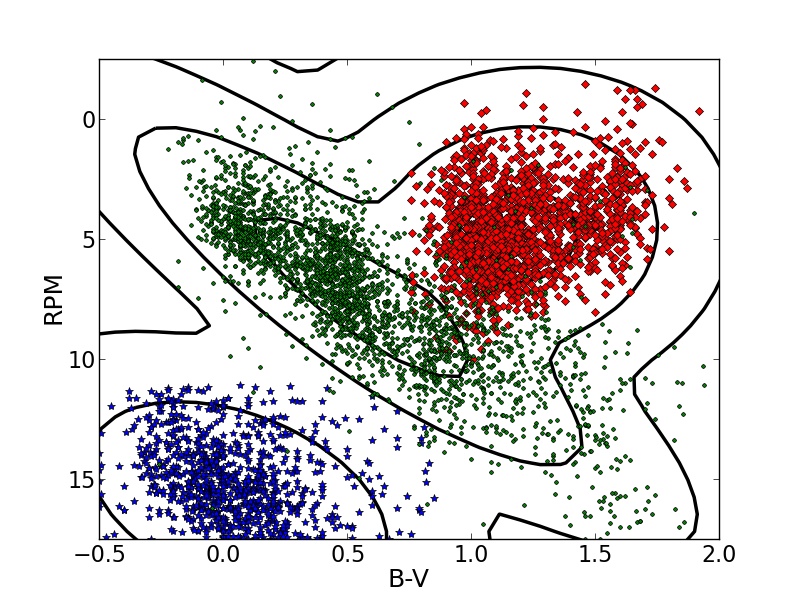}
\caption{The Evryscope Target Classification - We use B-V color differences and reduced proper motion (RPM) data with a two step machine learning algorithm to classify star size. \textit{Top:} the training data (gold squares=hot subdwarfs, grey=all others) for first step, the support vector machine (SVM) which returns the resulting hot subdwarf classification region (the area inside the black border). \textit{Bottom:} the training data (blue stars=white dwarfs, green=main sequence, red diamonds=giants) for second step, the Gaussian Mixture Model (GMM) which returns the resulting classification contours. Negative log likelihood plot-lines 1, 1.7, 2.8 are shown. This figure is originally presented in \cite{polarpaper} and reproduced here.}
\label{fig:classifier}
\end{figure}

In the first step of the machine learning process, we use a support vector machine (SVM) from the SKYLEARN python module \citep{scikit-learn} to identify likely hot subdwarfs (HSD) from all other stars. The HSD are challenging to separate since they can be close to main sequence B or A stars in this parameter space. We find the SVM to be an effective way to segregate the HSD, shown in the top panel of Figure \ref{fig:classifier} as the small confined area enclosed in the black border. This is done by using a training set of HSD from \cite{2017OAst...26..164G} and other types of stars from SIMBAD \citep{2000A&AS..143....9W}, filtering the outliers, then computing the contour boundaries. The SVM method is a non-probabilistic two-class classifier that computes a hard boundary (decision boundary) by minimizing the distance (or margin) between the points closest to the boundary. As with any classifier there are missed targets and contaminants, and there are physical reasons the results can be skewed (reddening for example). Our goal in this step is to separate the most challenging class (the HSD) from all the other classes while providing a boundary with a reasonable contingency space to the nearby white dwarf and main sequence regions.

Once the HSD are identified, all remaining objects are classified using a Gaussian Mixture Model (GMM) \citep{scikit-learn} with three classes to identify white dwarfs, main sequence, and giants. Although not the focus of this work, the solutions to main sequence stars and white dwarfs provide boundaries that are necessary as a comparison check to the HSD boundary from the first machine learning step described in the previous paragraph. We briefly describe the process here and refer the reader to \cite{polarpaper} for further details. The GMM method is a best fit to 2-D Gaussian function (probability density function), using the training points (20,972 main sequence, 1515 white dwarfs (WD), and 10,000 giants selected from SIMBAD) to adjust the Gaussian centers, orientations, and elongations. The GMM classifier results are shown in the bottom panel of Figure \ref{fig:classifier}. The GMM produces contour lines with Negative-log-likelihood (NLL) values that can be converted ($LH=10^{-NLL}$) to give an estimate of the confidence level the data point belongs in the class.

We use the spectral type and temperature profiles in \cite{2013ApJS..208....9P} to derive a function (using 1-D interpolation) that uses available color differences to derive an estimate for spectral type. The multiple color differences are averaged to choose the closest spectral type and luminosity class. The code produces a function with RPM and color differences inputs and outputs the star size, star type, and NLL score for the GMM step. We used this to select potential HSDs from our input catalog, with the added requirement that the HSD also be apparent spectral type O or B. The added requirements help filter contaminants from main sequence A stars. Further details on the design, testing, and performance of the Evryscope Classifier can be found in \cite{polarpaper}.

\subsubsection{Evryscope GAIA Classifier} \label{section_gaia_class}

The Evryscope GAIA Classifier uses the GAIA-DR2 G-band absolute magnitude (corrected using only the GAIA-DR2 parallax) and the GAIA-DR2 B-R color. The same support vector machine and Gaussian mixture model machine-learning approach from the Evryscope Classifier (see \S~\ref{section_es_class}) is used to define the classification contours. The same training set from \cite{polarpaper} is again used, but with the GAIA-DR2 data to generate the G-band absolute magnitude and B-R color space. Here:

\begin{equation}
    G_{abs} = G + 5 \log(Parallax/1000) + 10
\end{equation}

\subsubsection{Classifier Results and Potential Targets} \label{section_class_results}

\begin{figure}[htb]
\includegraphics[width=1.0\columnwidth]{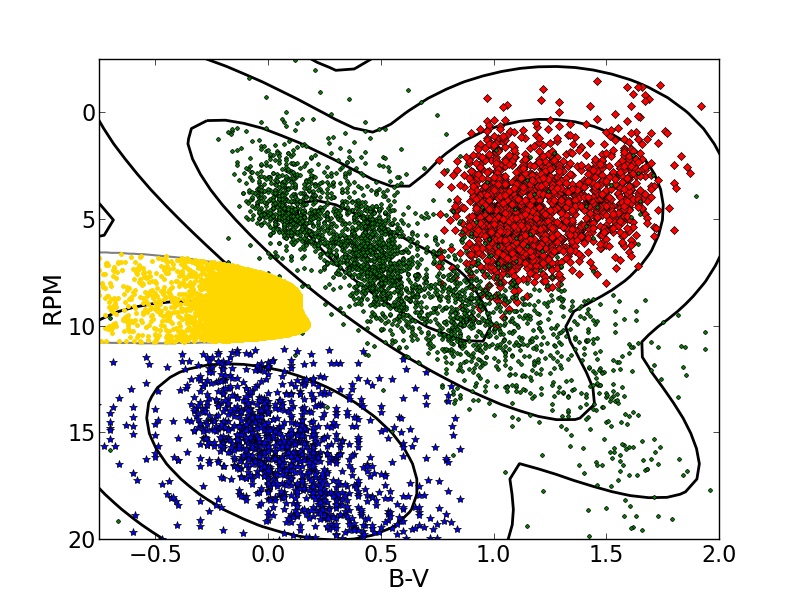}
\includegraphics[width=1.0\columnwidth]{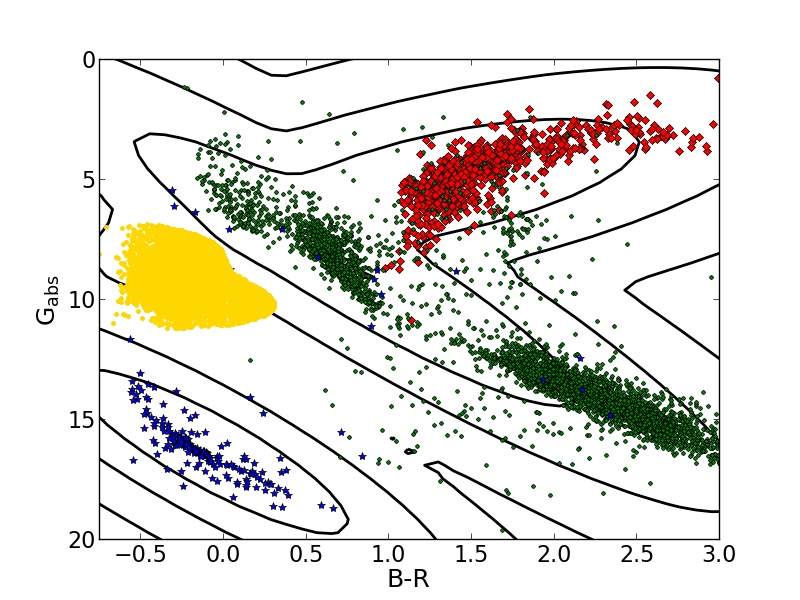}
\caption{The Evryscope Classifiers (see \S~\ref{section_es_class}), two step Machine Learning based spectral classifiers used to select HSD candidates. The black contours are the results of the GMM using training data from known giants (red diamonds), main sequence stars (green circles), white dwarfs (blue stars). The potential hot subdwarf (HSD) candidates are identified with a SVM step and are shown as the yellow grouping above the white dwarfs (WD) and to the left of the main sequence stars. \textit{Top:} The APASS / PPMXL based classifier. \textit{Bottom:} The GAIA-DR2 based classifier. We combine these results with external lists (\S~\ref{section_published_lists}) to identify objects as likely HSDs and check for photometric variability in the Evryscope light curves.}
\label{fig:classifier_gaia_fig}
\end{figure}

\begin{figure}[htb]
\includegraphics[width=0.95\columnwidth]{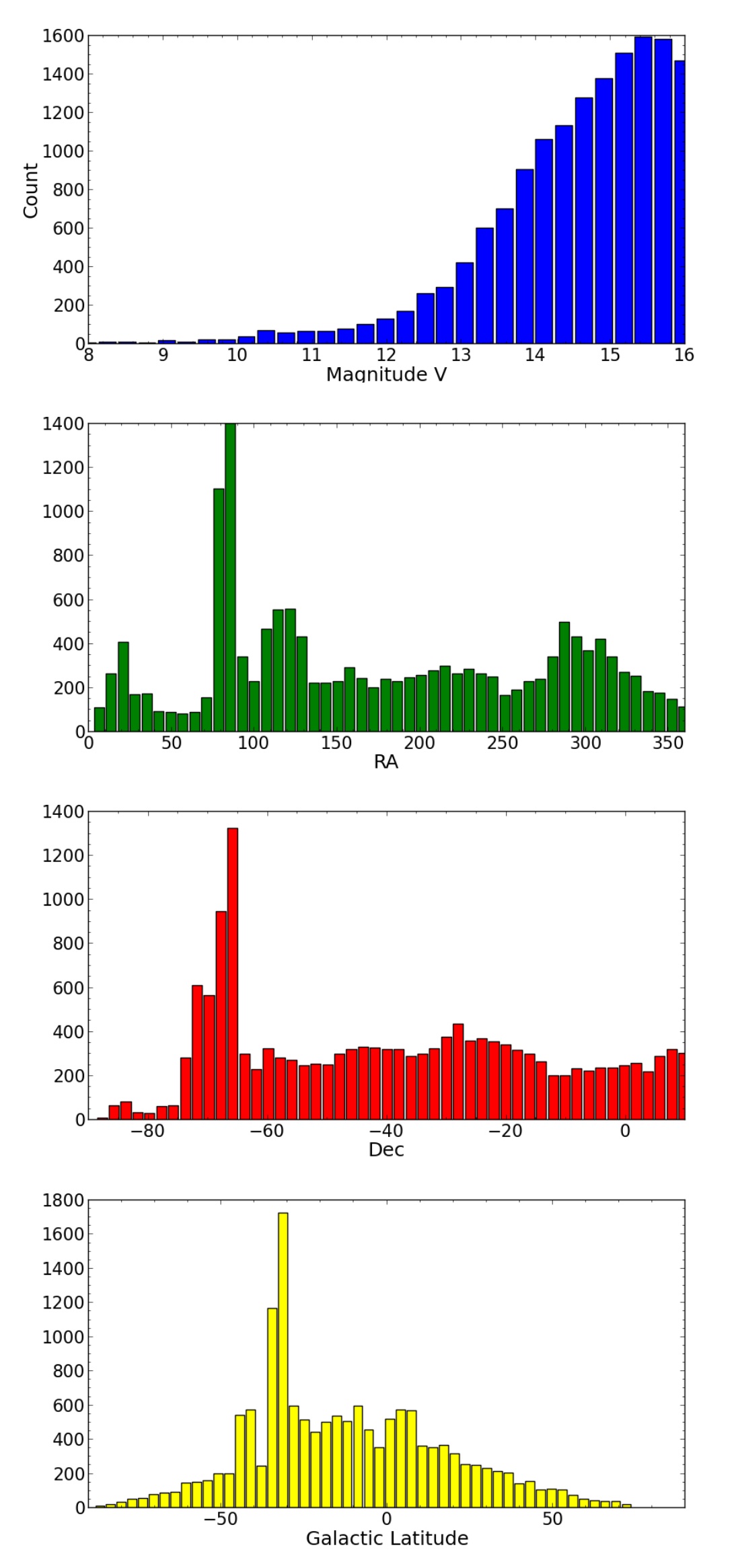}
\caption{The potential hot subdwarf (HSD) targets for the Evryscope survey. The distribution of targets in RA, Declination, and magnitude are as expected but with noticeable over-densities in the galactic plane and Large Magellanic Cloud (RA=80.89 Dec=-69.76). We apply an additional filtering step to flag likely impostor targets, biased toward not eliminating actual foreground HSDs that lie in these regions.}
\label{fig:targets_hist}
\end{figure}

The classifier results are shown in Figure \ref{fig:classifier_gaia_fig} with the HSD candidates in gold. We combine these results with external lists (\S~\ref{section_published_lists}) to identify objects as likely HSDs. Potential targets for the HSD survey is shown in Figure \ref{fig:targets_hist}; their distribution in RA, Dec, and magnitude are as expected. There are noticeable over-densities in the galactic plane and Large Magellanic Cloud (LMC). HSDs are not expected to be in the galactic plane or LMC at the bright magnitudes in our survey, however a significant number of viable targets should be visible in these fields as foreground stars. We use the results of the Evryscope database query to assist in identifying the HSDs in these challenging regions that are potentially useful to the survey.

\subsubsection{Light Curve Query} \label{section_lc_query}

The Evryscope database contains light curves for 9.3M targets with $m_{g} < 15$, with epochs through 2018. Dimmer sources and latter epochs are currently being processed. We added the additional filter requiring a minimum number of epochs (1000), and discarded sources with likely photometry issues, or likely crowding (indicated by excessive flags). The database query returned 11,220 potential HSD light curves from the 18,388 unique potential targets (which extend to $m_{v} =  16$) identified from the input lists described in the previous sections.

\subsubsection{Crowded Fields - Galactic Plane and LMC} \label{section_gp_lmc}

The crowded fields of the galactic plane and LMC are problematic due to blended sources, high background and increased noise, and the decreased accuracy of color difference measurements. Even with the GAIA-DR2 \citep{2018A&A...616A...1G} based data, \cite{2019A&A...621A..38G} found an excess density of targets in these fields and a higher rate of false positive HSDs. To address this issue they applied an additional filter in these regions based on excessive variance in the photometric measurements and background noise. \cite{2019A&A...621A..38G} also reported the distribution of distances for HSDs for a representative sample in their survey and show that $\approx$90\% of HSDs are within 2 kpc at a limiting magnitude of G=19. Given the galactic plane and LMC distances and the Evryscope HSD survey limit in this work of $m_{g} < 15$, we do not anticipate detecting any HSDs in these actual regions. However, we do expect some to be in the fields as foreground HSDs, such as AA Dor \citep{1978Obs....98..207K} and in densities similar to the rest of the survey. In general, sources in the galactic plane or LMC fields are expected to be problematic given the Evryscope pixel scale \citep{2015PASP..127..234L}. Evryscope targets in these regions are typically blended sources and the limiting magnitude increases by $\approx 1$ mag due to the increased background and other challenges.

To identify HSD impostors in crowded fields we compare the magnitude and coordinates from the input target lists to the values returned by the light curve query. This results in a magnitude error ($m_{g}$ or $m_{V}$ from the input lists less the mean Evryscope magnitude from the light curve) and an error in distance (the input list coordinates versus the centroid coordinates of the Evryscope pipeline). We also compare the list magnitude to the distance error. The average accuracy of correct targets is $\approx$ 1-2 arcsec and is clearly sub-pixel (13.3"). There is a significant bias toward recovering a brighter star indicating the likelihood of missed targets substituted with a nearby bright star (or blended with the nearby bright star so that the light curve is completely dominated by the bright source).

An analysis of the variable candidates from this work revealed that light curves returned from the target search that were more than 3 magnitudes in error or greater than 20 arcsec in distance were more than 90\% likely to be a wrong or blended target (referred to hereafter as blended). Applying these criteria to the survey query, 38\% of the HSD light curves showed signs of strong blending, with the dimmer sources suffering the greatest contamination. The majority of these blended targets (58\% of \textit{blended targets}) are sources in the galactic plane or LMC, expected to be problematic given the Evryscope pixel scale. Said another way, over $90\%$ of returned Evryscope light curves in the HSD survey in galactic plane or LMC regions are blended sources, consistent with the expectations in \cite{2015PASP..127..234L} and with the approximate factor of 10 in overdensity in potential HSDs of these regions in the target lists.

We also note here that the average blended targets rate for sources \textit{not} in the galactic plane or LMC is $\approx$ 16\%, in good agreement with predictions \citep{2015PASP..127..234L} for blended sources in the Evryscope images, given the Evryscope pixel scale.

\subsubsection{Blended Sources} \label{section_blended_sources}

Given the correlation between the likely blended targets identified in \S~\ref{section_gp_lmc} and the galactic plane or the LMC, the all sky blended source agreement with the predictions in \cite{2015PASP..127..234L}, and the agreement of overdensity in crowded regions to the blended sources in those regions, we conclude blended sources account for substantially all of the errant light curves and HSD impostors. To complete the survey in this work, two determinations must be made regarding the blended sources. First in regards to the inspection of the light curves and second in regards to estimating the number of HSD targets in the survey. 

The light curve query returned 11,220 potential HSD light curves (see \S~\ref{section_lc_query}), and later in the manuscript (\S~\ref{section_discussion}) we show the survey is target limited. Given the manageable number of light curves (each is visually inspected), and the need for targets, we inspect all light curves regardless of likely blended sources. A flag identifies if the source is in a problematic region or with large errors in distance or magnitude. In the event of a discovery, followup work reveals if the system is a HSD. Several discoveries later shown to be rare HSD systems (see \S~\ref{section_discovery}), were made in crowded fields that would have been missed had these targets instead been eliminated from the light curve query.

In estimating the number of HSD targets in the survey, we use the non-blended sources along with the recovery and false positive rates found in \S~\ref{section_list_perf}, to determine the likely number of HSDs. We then use the global blended sources rate (16\%) and analyze this much smaller group of likely HSDs (but blended with a nearby source) depending on the search type. Although any HSD transit signal would be greatly reduced from the blended brighter source, a fraction of the systems may detectable depending on how much brighter the contaminant is and how deep the variability signal. We discuss the contribution of these blended sources in \S~\ref{section_discussion}.

\subsubsection{Testing Spectral-ID Performance} \label{section_list_perf}

We tested the performance of the source lists in several ways, with the goals of quantifying the recovery rate and the false positive rate in order to estimate the likely HSD targets in our survey. We tested the targets to spectroscopically known HSDs from other works, and to confirmed HSDs from this work. The results are summarized in Table \ref{tab:classifier_sum_tab}.

(1) We compare the HSD targets from each list to the spectral type from the SIMBAD database \citep{2000A&AS..143....9W}. We require the coordinate crossmatch to be within 25 arcsec, and the magnitude comparison (GAIA $m_{g}$ or APASS $m_{V}$ vs SIMBAD $m_{V}$ where available) to be within 2 magnitudes. For crossmatched targets that have an available SIMBAD SpT (none or N/A are discarded), those with sdB or sdO matched to the lists are counted as recovered and the other spectral types are false positives. The recovery rates increase as the classification requirement is relaxed, however the false positive rate also increases (as expected and shown in Table \ref{tab:classifier_sum_tab}). The SIMBAD results show lower false positive rates for HSD's than the other testing methods (steps 3-4 following), however the comparative pattern between confidence levels is very consistent with the other testing methods. We attribute this difference to the less stringent SIMBAD classification of hot subdwarfs than that of the known HSD systems and spectroscopically confirmed HSD's.

(2)  We compare the HSD targets from each list to the spectroscopically verified known HSD's from \cite{2005A&A...430..223L, 2007A&A...462..269S, 2012MNRAS.427.2180N}. After filtering to our magnitude and declination range, the same crossmatch and magnitude comparison requirements from step (1) are used to identify which source lists recover the known targets. The recovery rates are shown in Table \ref{tab:classifier_sum_tab}. 

(3) The HSD survey recovered 79 known variables, described later in the manuscript in \S~\ref{section_discovery}. Fourteen of these are HSDs and the balance are variables of some other spectral class, the result of misclassification from one of the search lists. The most common contaminates were various variable types (RRlyrae, Cephied, Mira Cet, LPV, CV, and Novae), and the most common stellar contaminates were A and B stars. Of the 14 correctly classified known HSD variables, 9 were from 3 or 4 of the source lists (including both GAIA based lists), 1 originated from two source lists, and 4 appeared on a single list. The recovery rate for the 3 or 4 source based targets and the targets with both GAIA lists is 64.3\% (9/14).

Of the 65 misclassified known recoveries, none are classified on 3 or 4 of the lists. Five are classified with both GAIA based lists, and the remaining 59 are only classified from a single list (15 from the Geier GAIA list, 17 from the ES GAIA list, and the rest from the APASS/PPMXL based list). The false positive rate for the 3 or 4 source based targets is 0\%, 35.7\% (5/14) for targets appearing in both GAIA based lists, while individual lists show a false positive rate of greater than 60\%.

(4) Select HSD variability discoveries from this work (see \S~\ref{section_det_of_var}) are spectroscopically confirmed by ID spectra taken with the SOAR 4.1 m telescope at Cerro Pachon, Chile with the Goodman spectrograph \citep{2004SPIE.5492..331C}. The results of the classification from the spectra are compared to each of the source lists. The spectra provide a wavelength coverage of 3700-6000 \AA\ with a resolution of 4.3 \AA. The prominent hydrogen and helium features are easily identified and measured, along with the temperature from the continuum. Full details of our instrument setup and processing pipeline are provided in \citep{polarpaper}.

We obtained ID spectra for 36 of the discoveries, 12 are confirmed HSDs, and 24 are not HSDs (mostly main sequence B stars). Of the 12 correctly classified known HSD variables, 9 are from 3 or 4 of the source lists, and 10 are from both GAIA based lists; 11 are from the \cite{2019A&A...621A..38G} GAIA based list, and 10 are from the Evryscope GAIA classifier. The recovery rates for the 3 or 4 source based targets and the targets with both GAIA lists are 75.0\% (9/12) and 83.3\% (10/12). Targets from a single GAIA list return 91.7\% (11/12) and 83.3\% (10/12). The other lists show less return that the GAIA based lists; similar to the test of spectroscopically confirmed targets and known recoveries described in the previous paragraphs.

Of the 24 misclassified targets, one is classified on 3 or 4 of the lists and 4 with both GAIA based lists. The individual GAIA based lists have 13 and 14 misclassifications. The false positive rate for the 3 or 4 source based targets and for targets from both GAIA based lists is 10\% and 28.6\%. The false positive rate for the GAIA based list and the Evryscope GAIA classifier are 54.2\% (13/24) and 58.3\% (14/24).

The target selection performance is summarized in Table \ref{tab:classifier_sum_tab}.

\begin{table*}[htb]
\caption{Testing Target List Performance in Identifying HSDs}
\centering
\begin{tabular}{ l c c c c c c}
Comparison Test & Confidence Level & List Requirement & Recovery & Rate & False Positive & Rate\\
\hline
(1) SIMBAD SpT (sdB/sdO) & Very High & 3 or 4 source lists & 845/1199 & 70.5\% & 39/884 & 4.4\%\\
     & High & Both GAIA based lists & 988/1199 & 82.4\% & 100/1088 & 9.2\%\\
     & Medium & Geier GAIA List & 1155/1199 & 96.3\% & 381/1536 & 24.8\%\\
     & Medium & Evryscope GAIA List & 1012/1199 & 84.4\% & 352/1364 & 25.8\%\\
     & Global & All targets & 1199 & -- & 1639/2838 & 57.8\%\\
\hline
(2) Spectroscopically Known HSD's & Very High & 3 or 4 source lists & 97/140 & 69.3\% & -- & --\\
     & High & Both GAIA based lists & 123/140 & 87.9\% & -- & --\\
     & Medium & Geier GAIA List & 136/140 & 97.1\% & -- & --\\
     & Medium & Evryscope GAIA List & 126/140 & 90.0\% & -- & --\\
     & Global & All targets & 140 & -- & -- & --\\
\hline
(3) Recoveries of known variables\\
\hspace{5mm} (this work) & Very High & 3 or 4 source lists & 9/14 & 64.3\% & 0/9 & 0\%\\
     & High & Both GAIA based lists & 9/14 & 64.3\% & 5/14 & 35.7\%\\
     & Medium & Geier GAIA List & 9/14 & 64.3\% & 15/24 & 62.5\%\\
     & Medium & Evryscope GAIA List & 9/14 & 64.3\% & 17/26 & 65.4\%\\
     & Global & All targets & 14 & -- & 65/79 & 82.3\%\\
\hline
(4) Spectroscopically Confirmed\\
\hspace{5mm} (this work) & Very High & 3 or 4 source lists & 9/12 & 75.0\% & 1/10 & 10.0\%\\
     & High & Both GAIA based lists & 10/12 & 83.3\% & 4/14 & 28.6\%\\
     & Medium & Geier GAIA List & 11/12 & 91.7\% & 13/23 & 54.2\%\\
     & Medium & Evryscope GAIA List & 10/12 & 83.3\% & 14/24 & 58.3\%\\
     & Global & All targets & 12 & -- & 24/36 & 66.7\%\\
\hline
    Test Summary (Averaged Values) & Very High & 3 or 4 source lists & -- & 70\% & -- & 5\%\\
     & High & Both GAIA based lists & -- & 79\% & -- & 24\%\\
     & Medium & Geier GAIA List & -- & 87\% & -- & 47\%\\
     & Medium & Evryscope GAIA List & -- & 81\% & -- & 50\%\\
     & Global & All targets & -- & all & -- & 69\%\\
\hline
\multicolumn{7}{l}{See \S~\ref{section_list_perf} for further details.}\\

\end{tabular}
\label{tab:classifier_sum_tab}
\end{table*}

\subsubsection{Other Considerations} 

The primary challenge of selecting the targets (common to all of the methods), is balancing the missed targets with the false positives. The \cite{2019A&A...621A..38G} GAIA based list very effectively selects HSD candidates with a fraction of contaminants; they estimate the primary contamination is from cool stars, blue horizontal branch, and post-AGB stars. We measure the false positive rate to be 47\% (see Table \ref{tab:classifier_sum_tab}) in our FoV and magnitude range, reasonable given the difficulty in separating HSDs from impostors. The Evryscope GAIA classifier uses an alternate color space and different selection approach to include some extra candidates and exclude others. The Evryscope APASS / PPMXL based classifier has a higher contamination rate, but includes more potential targets and it is the same source catalog we use for our forced photometry pipeline. The HSD survey in this work is target limited, and benefit from the extra potential targets. Additionally, we developed a few ways to segregate the likely targets from the impostors. 

A powerful feature of our target selection method is the duplication of sources. Based on the list performance false positive results discussed above and shown in Table \ref{tab:classifier_sum_tab}, candidates identified in 3 or 4 of the lists are greater than 90\% likely to be HSDs. Candidates with both GAIA based sources are greater than 70\% likely to be HSDs. Using this along with the recovery rates, we identify the high-likelihood targets and estimate the total number of HSDs in our search.

The compact nature of HSDs drives the transit and eclipse times. They are expected to be fast ($\approx 20$ minutes) with deep depths. A light curve from an eclipsing binary with a main sequence A star would have a much longer (3-4 hour) eclipse time indicating the target is likely to be a HSD imposter, even more so if the classifier results are marginal. Eclipsing binary candidates in this work with marginal classifier results and long eclipse durations were identified as low-priority followup (given the HSD focus of our search), and presented in the appendix.

\subsubsection{Summary of Targets} \label{section_sum_tar}

To estimate the likely number of targets from each survey, we begin with the number of light curves returned from the database query for each classifier confidence level. The totals are adjusted by the likely blended targets fraction. From the rates in Table \ref{tab:classifier_sum_tab}, we calculate the average recovery and false positive values per confidence level. We divide the adjusted number of targets by the recovery rate and subtract the false positives to estimate the total sources. A summary of the HSD targets is shown in Table \ref{tab:targets_tab}.

\begin{table}[h]
\caption{Survey Targets}
\begin{tabular}{ l c c c c}
\textbf{HSD}\\
Total Targets & Confidence & Recovery & False Positive & Likely HSDs\\
\hline
1071 & Very High$^a$ & 70\% & 5\% & 1203\\
1843 & High$^b$ & 79\% & 24\% & 1087\\
3497 & Medium$^c$ & 87\% & 47\% & 1314\\
3465 & Medium$^d$ & 81\% & 50\% & 1341\\
11,220 & Global & all & 69\% & 2167\\
\hline
Survey & & & & 1422 $\pm{428}$\\
\hline
\multicolumn{5}{l}{$^a$Requires targets selected from 3 or 4 source lists (see \S~\ref{section_list_perf}).}\\
\multicolumn{5}{l}{  Note: The very high confidence level likely HSD number}\\
\multicolumn{5}{l}{  is extrapolated, since the false positive rate is much lower}\\
\multicolumn{5}{l}{  than the missed targets (1-recovery) rate.}\\
\multicolumn{5}{l}{$^b$Requires targets selected from both GAIA based source lists.}\\
\multicolumn{5}{l}{$^c$Requires targets selected from the Geier GAIA based source list.}\\
\multicolumn{5}{l}{$^d$Requires targets selected from the ES GAIA based source list.}\\
\multicolumn{5}{l}{See \S~\ref{section_sum_tar} for calculation of Likely HSD's}\\
\end{tabular}
\label{tab:targets_tab}
\end{table}

\subsection{HSD frequency}

The Evryscope database contains 9.3 million light curves for stars brighter than 15.0M in $m_{g}$, and we estimated that 1422 of these are hot subdwarf stars. The HSD frequency is $1422 / 9.3M$ or $\approx$ 1 in 10,000 stars in the Evryscope field are HSDs. A space-density conversion is beyond the scope of this paper as the primary goal here is searching for photometric variation. In \S~\ref{section_HSD_survey_2} we discuss the followup HSD survey (Evryscope HSD survey 2) to this work, which will be expanded in FoV (North and South all-sky coverage), limiting magnitude, and observational coverage. We will estimate the space-density in the Evryscope HSD survey 2.

\subsubsection{Survey Completeness}

From the classifier testing described in \S~\ref{section_list_perf} (methods (1) and (2)) we compare the total HSD recovered to the total available in our magnitude and declination range. This leads to SIMBAD completeness rate of 80\% (1199 / 1499) for the HSD survey. The confirmed spectra completeness rate is 86\% (140 / 162). We average these rates and decrease the results by the likely wrong target fractions from the previous section to estimate the completeness of 51\% for southern sky targets brighter than 15.0M in $m_{g}$ for the HSD survey.

\section{Detection of Variables} \label{section_det_of_var}

\subsection{Detection Process}

All of the 11,220 potential HSD light curves were visually inspected for variability. The classification confidence level and distance flags are included for each source to help evaluate the likelihood the target is a HSD and to prioritize followup. Prior to inspection, the light curves were first processed to remove systematics and identify nearby reference stars for comparison as described below.

The timestamps in the Evryscope light curves were converted from Modified Julian dates to Heliocentric Julian dates using PyAstronomy's \textit{helCorr} function. We pre-filtered the light curves with a Gaussian smoother to remove variations on periods greater than 30 days, and a 3rd order polynomial fit was subtracted to remove long-term variations. Light curves were then searched for transit-like, eclipse-like, sinusoidal and quasi-sinusoidal variability signals using the Box Least Squares (BLS) \citep{Kovacs:2002gn, 2014A&A...561A.138O} and Lomb-Scargle (LS) \citep{1975Ap&SS..39..447L, 1982Ap&SS..263..835S} algorithms, along with a custom search tool (the outlier detector, see \S~\ref{section_outlier_alg}) designed to find fast transits characteristic of HSDs. Details of the algorithm settings are described in the following sections.

The target light curve (both folded and unfolded) is compared to nearby reference stars for any indications that the detected signals may be systematics. The plots are colored by time to check how well-mixed the detection is, since a transit or eclipse with only a few occurrences is more likely to be an artifact of the detection algorithm. All detections are filtered to mask likely daily-alias periods indicative of systematics as described in \cite{polarpaper}. The power spectrum of each detection algorithm is displayed for each target, however we do not filter by power. All targets are visually inspected, as we wish to search potential candidates since lower detection signals may be indicative of shallow transits or fast transits with only a few periods captured. Variability candidates were then vetted with a separate reviewer confirming the candidate light curves.

\subsection{Variability Search Algorithms}

\subsubsection{Conventional Search Algorithms} \label{section_conv_alg}

\begin{figure}[htb]
\includegraphics[width=1.0\columnwidth]{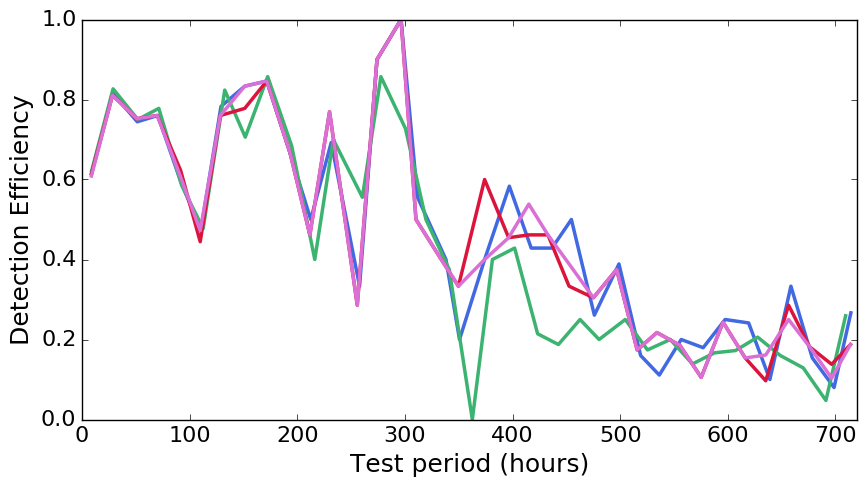}
\caption{Detection efficiency of known variables in the FOV, magnitude, and amplitude ranges of the HSD survey with different BLS and LS settings. Green line: BLS maximum period 240 hours, number of periods 25,000, and LS maximum period of 720 hours. Blue line: BLS maximum period 480 hours, number of periods 50,000, and LS maximum period of 1440 hours. The red and magenta lines hold the same long period BLS and LS settings, but with coarse period sampling shown in the red (25,000) and finer period sampling shown in the magenta (100,000). These tests on known variables helped establish the transit fraction and number of periods in order to effectively cover the period search range of 2-720 hours. We used simulated transits in \S~\ref{section_alg_perf} to confirm the final settings.}
\label{fig:det_settings}
\end{figure}

We tested different BLS settings to maximize the recovery rates on Evryscope light curves in \cite{polarpaper}, a variability survey of the southern polar region (Evryscope polar search). The HSD survey features longer light curve coverage (2.5 years compared to 6 months in the polar search), and the variability signals are expected to be faster. Consequently, we extended the period coverage and retested the settings on known variables in our magnitude range, with amplitudes we expected to find in the HSD surveys (0.01 to 0.50 in fractional normalized intensity). We verified the setting adjustments did not hinder detection performance at the shorter periods, as demonstrated in Figure \ref{fig:det_settings}. The final BLS settings used on the HSD search were a period range 2-480 hours with 50,000 periods tested and a transit fraction of 0.01 to 0.25. We used an LS range of 2-720 hours.

The lower period cutoff in the BLS and LS period range does not preclude us from finding one hour signals or less as aliases of the true period, demonstrated by our recovery of the known 1.17 hour system CD-30 11223 shown later in the manuscript (see \S~\ref{section_discovery}). Very short period binaries (with tens-of-minutes periods) would benefit from additional systematics processing and modified detection algorithms. We discuss these modifications and the potential very short period search in \S~\ref{section_compact_binary_discussion}.

\subsubsection{The Outlier Custom Search Algorithm} \label{section_outlier_alg}

HSD transits are expected to be fast (on the order of tens-of-minutes), and deep (up to completely eclipsing if the orientation is optimal, e.g.: Konkoly J064029.1+385652.2 \citealt{2015ApJ...808..179D}). Even with periods as short as a few hours, the transit fraction is still small, and the most significant points in the light curve are very dim outliers. This situation is quite different than the traditional shallow (less than 1\%) and longer (at least a few hours) transits BLS was designed to find, and completely different than the sinusoidal signals LS excels at. We developed a custom code, called the outlier detector, to find the narrow and deep signals characteristic of HSD transits. Although not the focus of this survey, transits of white dwarfs are expected to be even faster (on the order of a few minutes) than HSD transits and also very deep. The outlier detector was developed to find both HSD and WD transit signals, given the similarities. The results of our WD transit survey will be discussed in an upcoming work (Ratzloff et al., in prep).

The outlier detector uses several iterative approaches to search for fast transits. The light curve is normalized (in flux), then the 1-$\sigma$ error is computed. Data points with a normalized flux value of 3-$\sigma$ below the mean are flagged. The number of flagged points is compared to a minimum value (set by the survey type, periods searched, and expected variability). For the HSD search, the minimum value is 50 (determined by requiring at least 5 transits with each 20 minute transit consisting of 10 data points given the Evryscope's 2-minute cadence). If the number of flagged points is less than the minimum value, the processed is restarted using 2.9-$\sigma$ and continues with .1 reductions in the $\sigma$ requirement until the minimum number of flagged points is met. In almost all cases where there is an actual fast, deep transit (from astrophysical or simulated signals) the original 3-$\sigma$ cutoff selects many more points than the minimum value given the tens of thousands of epochs in the typical Evryscope light curve. This initial iterative process helps the limiting case of a long period, fast transit that may only have a few transits even in a multi-year light curve.

The flagged points (i.e. the outlier points) are then kept and all other points discarded for the next steps. The outlier points are then phase folded at 250,000 different periods (spread evenly in period space) in the test period range. For both the HSD and WD searches, we tested periods from 2-480 hours. For each period, we calculate the standard deviation in \textit{phased-time}, without regard to the normalized flux of the outlier points. The first step (described in the previous paragraph) set the outlier points, here we are only interested in how well the points align in phased-time. We then sigma-clip the outlier points (using 3 iterations and 2-$\sigma$ from the mean in phased-time). This sigma-clip step helps remove errant low flux points not associated with the periodic signal. We recalculate the standard deviation in \textit{phased-time} of the sigma-clipped outlier points. The period with the lowest standard deviation, calculated in this way, is selected as the best period.

The same process is repeated for a smaller range ($\pm$ 3 minutes centered on the best period), testing 5000 periods, but in finer increments than in the previous step. This fine period step narrows the detection period, and increases the accuracy to levels necessary in very short period HSD and WD systems.

\begin{figure}[htb]
\includegraphics[width=1.0\columnwidth]{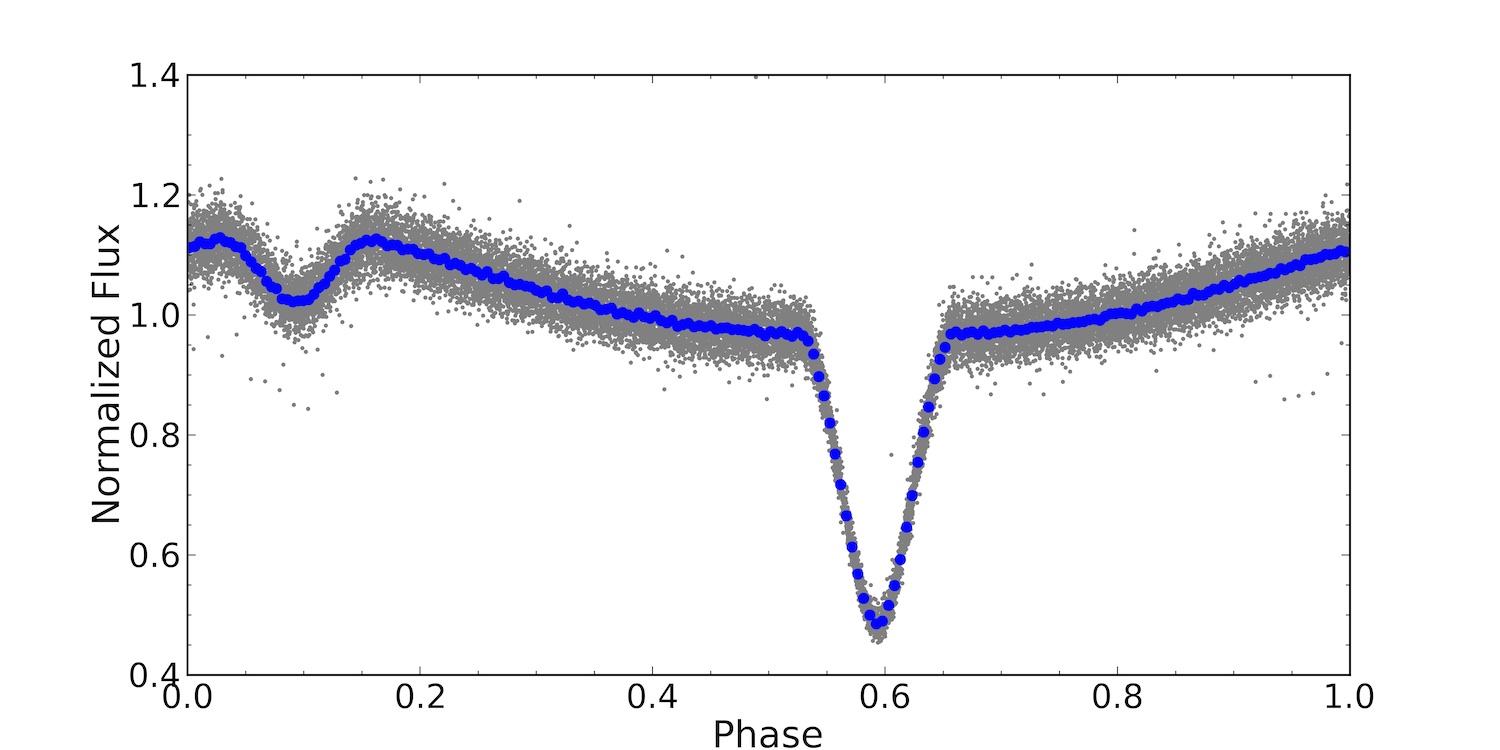}
\includegraphics[width=1.0\columnwidth]{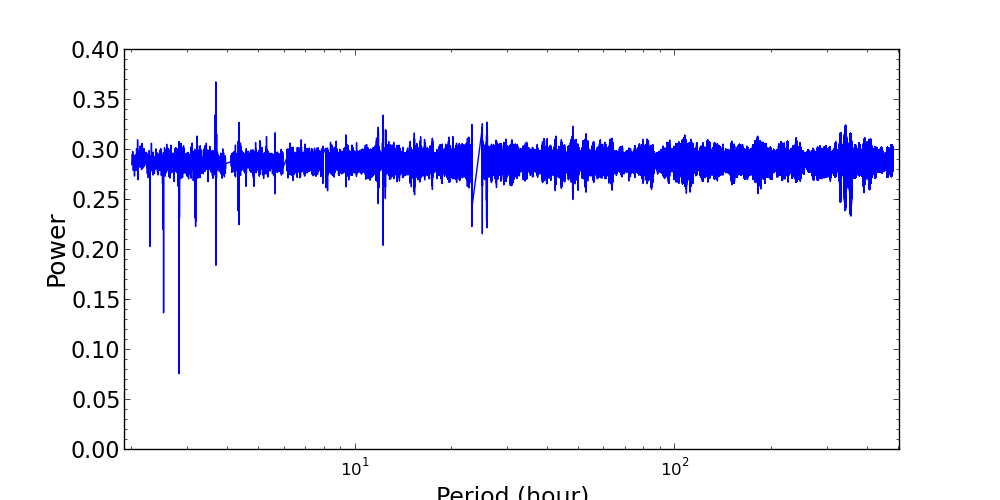}
\caption{\textit{Top:} The Evryscope light curve of the known HSD system HW Vir folded on its period of 2.80126 hours. Grey points = 2 minute cadence, blue points = binned in phase. \textit{Bottom:} The outlier detector (the fast-transit algorithm \S~\ref{section_outlier_alg} designed for the HSD and WD surveys) power spectrum with the minimum spike at the 2.80126 hour detection.}
\label{fig:det_outlier}
\end{figure}

An example detection from outlier detector for the known short period HSD system HW Vir is shown in Figure \ref{fig:det_outlier}. The best detection is the minimum spike (the period phase-folded with lowest standard deviation of the outlier points) at the 2.80126 hour period. Here, the deep transit ($\approx$ 0.50) of the primary drives the detection, the variation from the secondary or the reflection effect is inconsequential. 

The outlier detector uses only a subset of points significantly below the mean flux value as it tests the best fit over the period range. This lessens the processing burden since the many fold periods are tested with a much smaller number of data points. The outlier detector code is optimized for speed and takes $\approx$ 5 seconds to run on an average Evryscope light curve (similar to BLS and LS). Expressing the power in terms of the standard deviation in phased-time has the added benefit of stabilizing the test space - periods with no particular signal cluster around .3 since the range is between 0 and 1 and in this situation the mean is .5. Alias periods tend to be suppressed or score poorly since the standard deviation of the outlier points in this situation is increased. In light curves with deep, fast transits the outlier detection signal tends to be sharp and separated from the noise floor.

\subsection{Search Algorithm Performance} \label{section_alg_perf}

We use 150 Evryscope light curves, distributed in magnitude, RA, and Dec, as the basis for testing our search algorithm performance. Since the planet transit signals are the most difficult to recover, we developed the custom search algorithm to find the fast and deep planet transits. The HW Vir type eclipsing binaries are recovered more easily due to the larger companion. HSD transit signals are injected onto the Evryscope light curves for a variety of planet sizes and periods. To create the transit signals, we assume a uniform source and use the analytical solution from \cite{2002ApJ...580L.171M} to generate a transit curve for each planet size and period. We use 250 points per transit (which translates to $\approx$ 5 seconds in time between points and varies slightly depending on the period) to ensure a fine sampling that captures the ingress, transit, and egress features. We then repeat the generated transit curve to cover the complete time coverage of the Evryscope light curves. The generated transit curve points are then averaged in groups to match the Evryscope integration time 2 minutes, typically with $\approx$ 25 points averaged to simulate a 2 minute epoch. We choose a random point in the first period of the generated transit curve and assign it to the Evryscope timestamp; all other times are propagated from this initial epoch. Matching times in the generated transit curves and the Evryscope light curves are then multiplied (in normalized flux) so that any transit values are injected into the Evryscope light curves, while preserving variation from the actual light curves.

The HSD simulations assume a star size of 0.2 $R_{\odot}$. For each planet size, the period search range is split into 100 test periods. We perform 15 iterations at each test period (each iteration adjusted with a random variation  of $\pm$ 1 \%), for each light curve, and test if the transit is detected. A detection is counted if either BLS, LS, or the outlier detector finds the period or an alias (half, 1/3, 1/4 or double, triple, or quadruple the period) within 1\% of the correct period.

\begin{figure}[htb]
\includegraphics[width=1.0\columnwidth]{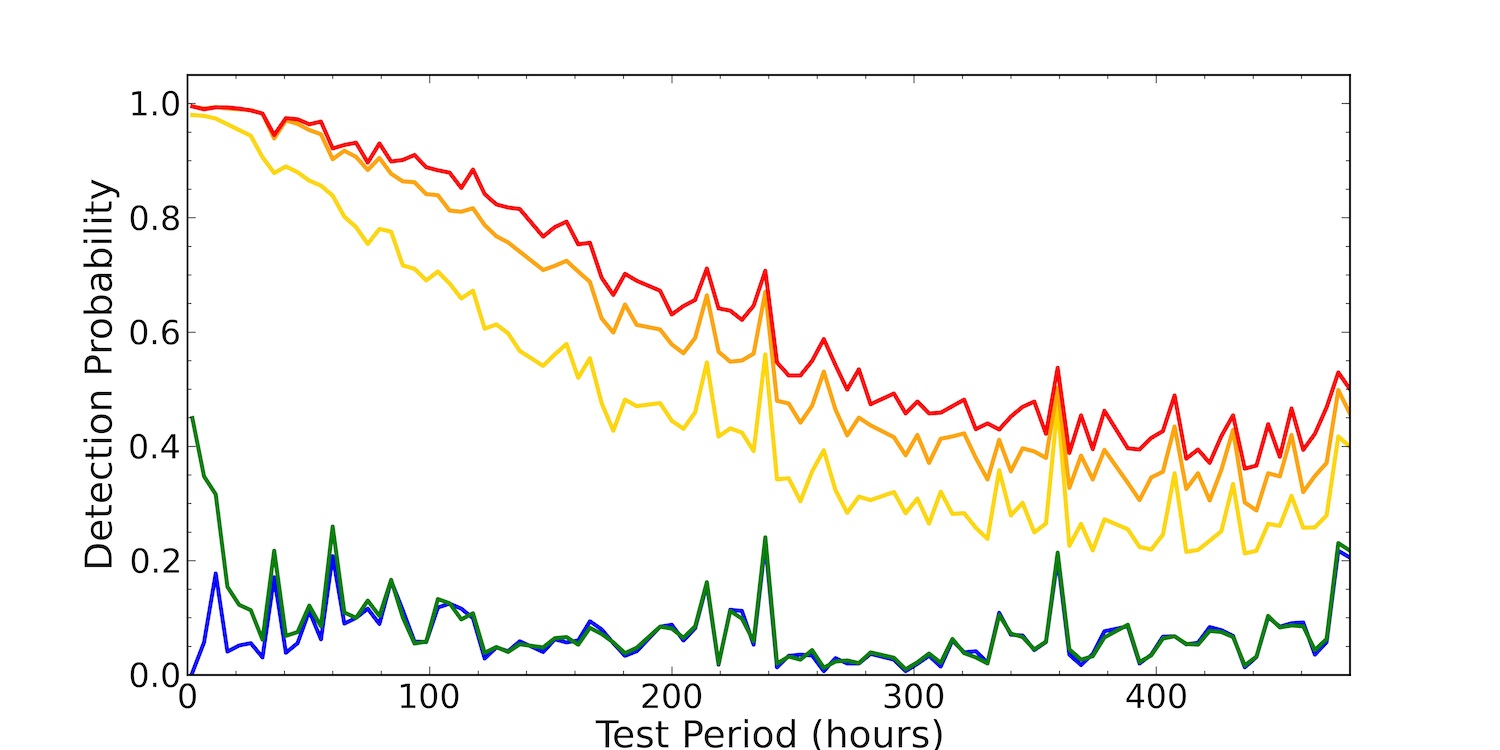}
\caption{The simulated recovery of HSD transiting planets with the Evryscope light curves and detection algorithms. The simulated transits are shown in decreasing size from red to blue. Red = late M-dwarf or brown-dwarf (.15 $R_{\odot}$), orange = Super-Jupiter (.125 $R_{\odot}$), yellow = Jupiter (.1 $R_{\odot}$), green = Neptune (.035 $R_{\odot}$), blue = Earth (.01 $R_{\odot}$). The simulation results here assume an inclination angle of $i=90^\circ$. In \S~\ref{section_discussion}, we calculate the transit fraction and survey sensitivity per planet size.}
\label{fig:det_sims_plots}
\end{figure}

We used 5 planet sizes to test the HSD recovery rates, ranging from Earth to Super-Jupiter size, with 1.1M simulations performed. Each simulation takes approximately 25 seconds, requiring 7500 hours of computing time, which we performed with a 16 core / 32 thread machine with an Intel(R) Xeon(R) Gold 6130 CPU @2.10GHz and 128 GB of DDR4-2666 RAM. From these tests, we expect high sensitivity to gas giant planets with periods up to at least 250 hours, with decreasing recovery at longer periods and smaller planets. The detection efficiency tests are shown in Figure \ref{fig:det_sims_plots}. We note the detection floor is near a Neptune size planet for all but the shortest periods. The simulation results here assume an inclination angle of $i=90^\circ$, and are the maximum expected recovery values. Further in the manuscript in \S~\ref{section_discussion}, we calculate the transit fraction and propagate the final detection probabilities and survey sensitivities per planet size.

\subsection{False Positive Tests} \label{section_fpt}

To test the discovery candidates, we compare the target light curve to the light curves from nearby reference stars looking for signs of similar variation indicative of systematics. We also check how well mixed in phase the detected period is, looking for data gaps or poor mixing that might be a result of the matched-filter fitting to systematics instead of an astrophysical signal. A separate researcher reviewed all the discovery candidates, and those with suspect quality or detection were thrown out. Additional details of the false positive tests performed by the Evryscope lab are available in \cite{polarpaper}. All discoveries in this work were folded on alias periods (half and double of the detected period) looking for additional signs of systematics, and to verify the correct period.


\section{FOLLOWUP OBSERVATIONS AND ANALYSIS} \label{section_followup}

\subsection{SOAR / Goodman ID Spectroscopy}

We obtained spectra for select HSD variability candidates on February 9, 2019, March 5, 2019, August 2, 2019, and September 9, 2019 with the Goodman spectrograph \citep{2004SPIE.5492..331C} on the SOAR 4.1 m telescope at Cerro Pachon, Chile. We use the 600 mm$^{-1}$ grating blue preset mode, 2x2 binning, and the 1" slit. This configuration provided a wavelength coverage of 3500-6000 \AA\ with a spectral resolution of 4.3 \AA\ (R$\sim$1150 at 5000 \AA). We took four 360 s spectra of all targets and the spectrophotometric standard star BPM 16274. For calibrations, we obtained 3 x 60 s FeAr lamps, 10 internal quartz flats using 50\% quartz power and 30 s integrations, and 10 bias frames.

\begin{figure*}[htb]
\includegraphics[width=2.0\columnwidth]{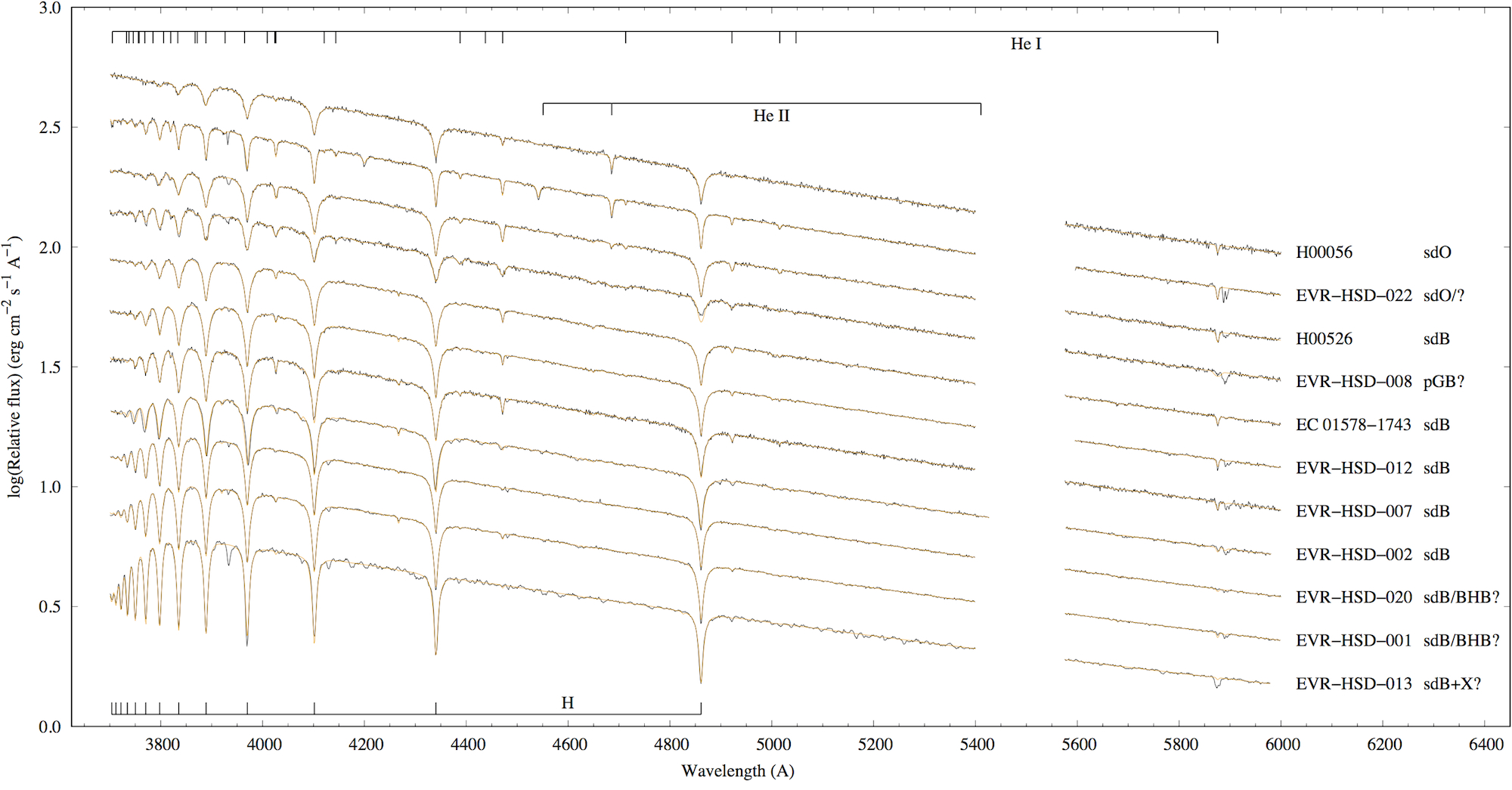}
\caption{Subluminous stars (black) in the {\sc Everyscope} sample together with their best-fit {\sc Tlusty/XTgrid} models (orange). The sample covers a wide range of objects along the blue horizontal branch from 20,000 K to 45,000 K surface temperature and gravity  $\log{g}>4.6$\,cm\,s$^{-2}$. The observed continua have been adjusted to the models to improve the figure.}
\label{fig:spec_hsds}
\end{figure*}

\begin{figure*}[htb]
\includegraphics[width=2.0\columnwidth]{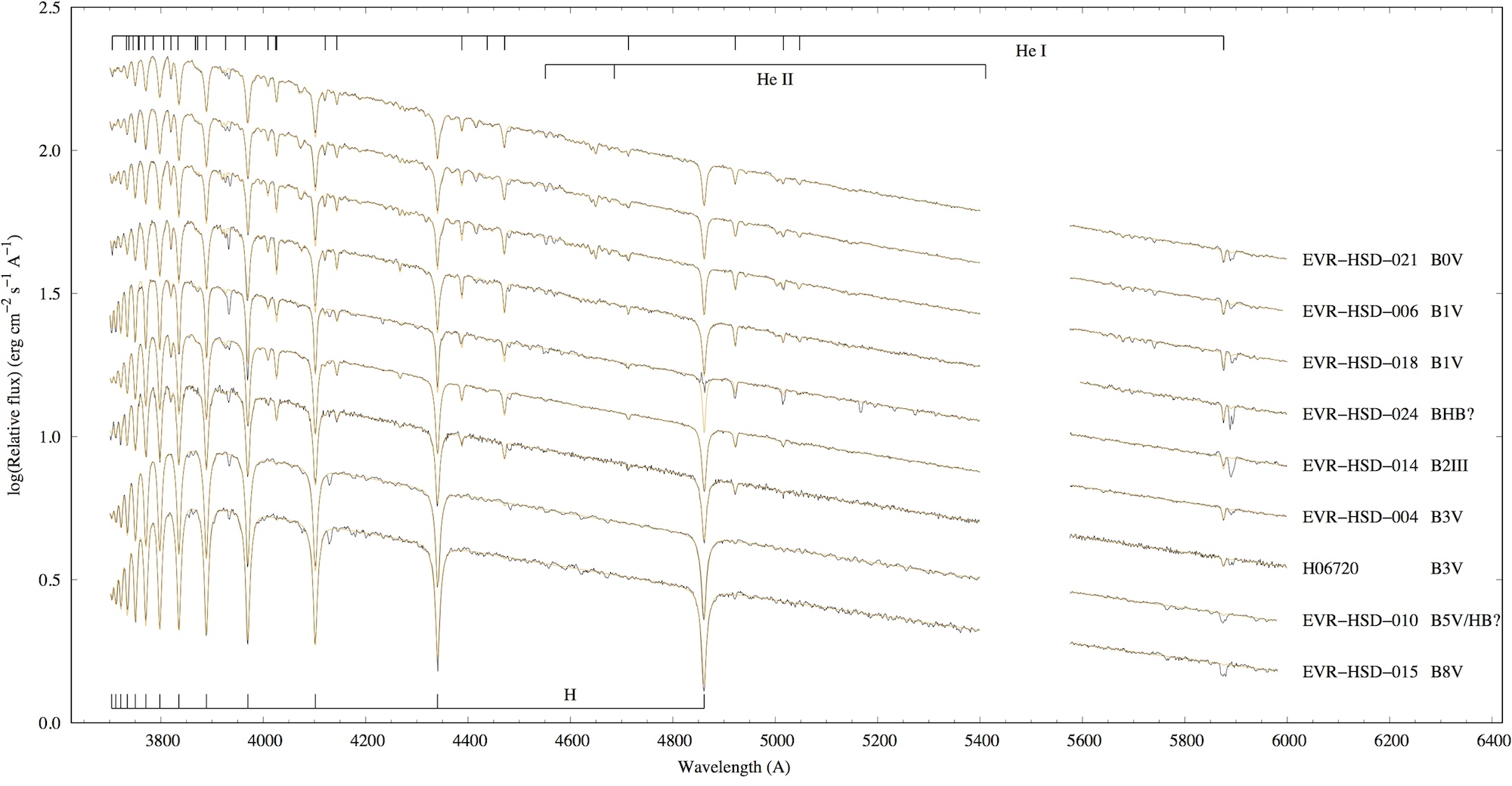}
\caption{Main sequence O and B type stars (black) in the {\sc Everyscope} sample together with their best-fit {\sc Tlusty/XTgrid} models (orange). The sample covers a wide range of objects from 12,000 K to 55,000 K surface temperature and gravity $\log{g}<4.5$\,cm\,s$^{-2}$. The observed continua have been adjusted to the models to improve the figure.}
\label{fig:spec_imposters}
\end{figure*}

\begin{figure*}[h]
\includegraphics[width=1.8\columnwidth]{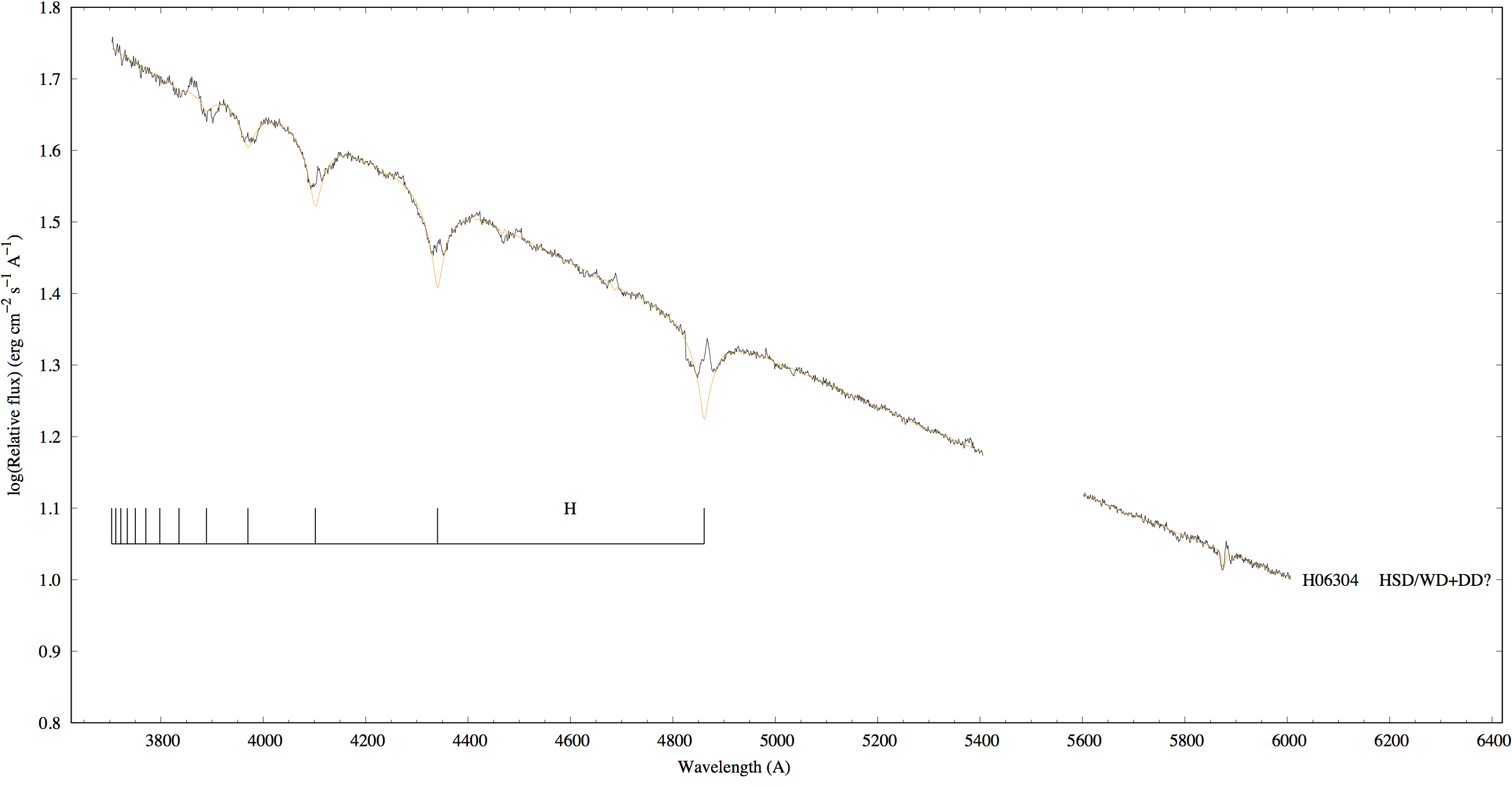}
\caption{A cataclysmic variable like spectrum together with a 40,000 K DAO type white dwarf model (orange). The observed continuum have been adjusted to the model to improve the figure.}
\label{fig:spec_debris_disc}
\end{figure*}

We processed the spectra with a custom pipeline written in Python; designed to extract, wavelength calibrate, and flux calibrate the spectra (optimized for this wavelength coverage and instrument setup). For additional details, we refer the reader to \cite{2019ApJ...883...51R}, where the pipeline is explained fully. We detect strong H Balmer lines in all variability candidates and He lines in many. Each spectrum was visually inspected and fitted using the stellar atmosphere model service for early type stars from \textit{Astroserver}\footnote{http://www.astroserver.org} \citep{2017OAst...26..179N}. From the best fits, we measure the effective temperature, surface gravity, projected rotational velocity, helium abundance and approximate the metallicity. For the metallicity, we use the C, N, and O abundances as a proxy. From these parameters we determine the spectral type. Discoveries determined to be HSDs are presented later in the manuscript in the top panel of Table \ref{tab:var_discoveries_sum}, and the false positives (main-sequence B stars in almost all cases) are shown in the the bottom panel of the same table. The spectra and best fits for the subluminous stars are shown in Figure \ref{fig:spec_hsds} and HSD imposters (mostly main sequence B stars) are shown in Figure \ref{fig:spec_imposters}. The spectrum for a potential debris disc is shown in Figure \ref{fig:spec_debris_disc}.

\subsubsection{ID Spectra Analysis with Astroserver}

The \textit{Astroserver} service uses {\sc XTgrid} \citep{2012MNRAS.427.2180N} which has been developed to automate the spectral analysis of early type stars with {\sc Tlusty/Synspec} \citep{2017arXiv170601859H, 2017arXiv170601935H, 2017arXiv170601937H} non-Local Thermodynamic Equilibrium (non-LTE) stellar atmosphere models. The procedure applies an iterative steepest-descent chi-square minimization method to fit observed data. It starts with a initial model and by successive approximations along the chi-square gradient it converges on the best fit. The models are shifted and compared to the observations by a piecewise normalization, which also reduces systematic effects, such as blaze function correction, or absolute flux inconsistencies due to vignetting or slit-loss. {\sc XTgrid} calculates the necessary {\sc Tlusty} atmosphere models and synthetic spectra on the fly and includes a recovery method to tolerate convergence failures, as well as to accelerate the converge on a solution with a small number of models.

During parameter determination of hot stars the completeness of the opacity sources included, and departures from LTE are both important for accuracy. We concluded that {\sc Tlusty} models with H, He, C, N and O composition deliver reliable results given the spectral resolution, coverage and signal-to-noise of the survey data. Although Mg and Si lines are visible in many spectra, these elements have relatively small effects on the atmospheric structure compared to C, N and O. 

\begin{figure*}[h]
\includegraphics[width=1.0\columnwidth]{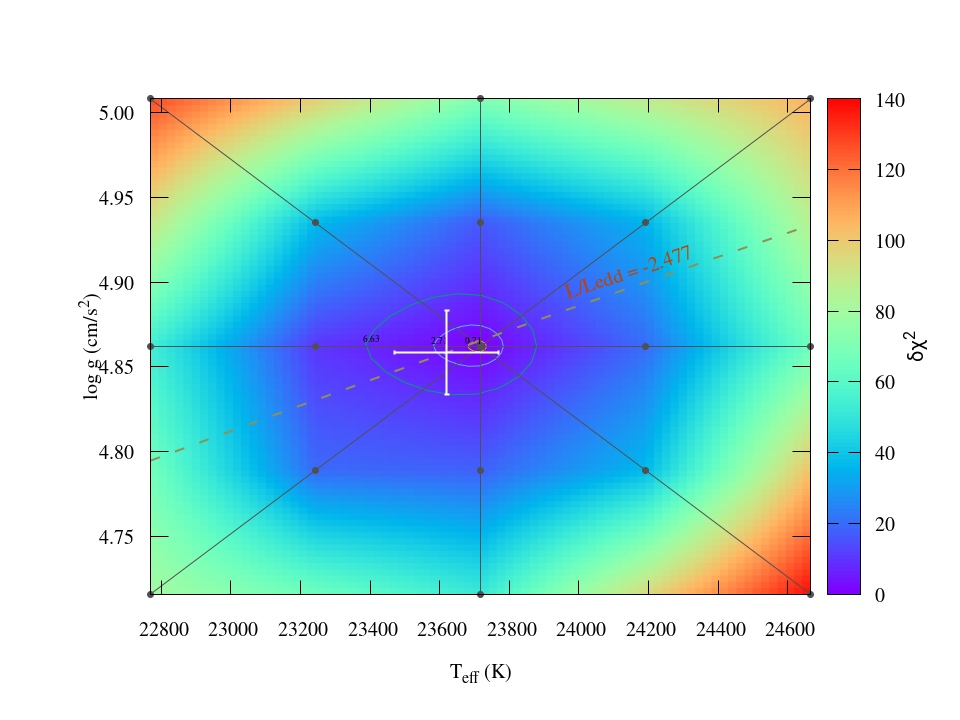}
\caption{Surface temperature and gravity correlations for EVR-HSD-020. The 40, 60 and 99\% confidence interval contours are marked. The white error bars show the final results. The dashed line is the iso-Eddington-luminosity curve corresponding to the best-fit.}
\label{fig:spec_conf_levels}
\end{figure*}

Parameter errors are evaluated by mapping the chi-square statistics around the solution. The parameters are changed in one dimension until the 60\% confidence limit is reached. Correlations near the best-fit values are also included in the final results as demonstrated for surface temperature and gravity for a representative example and shown in Figure \ref{fig:spec_conf_levels}.

\subsection{TESS Photometry}

HSD variable discoveries were confirmed with TESS light curves (where available) using the ELEANOR pipeline \citep{2019PASP..131i4502F}. We used the PSF photometric setting for bright stars ($m_{g} < 12.0$) and the standard aperture setting for all other targets. We also verified there was not a significant light curve variation between the different settings. The ELEANOR pipeline data are from TESS full frame images (FFI) with a 30 minute cadence. For one of the discoveries (EC 01578-1743 see \S~\ref{section_rb_ec015781743}) we instead used the available TESS TOI light curve, which has a 2 minute cadence. The TESS followup and Evryscope discovery light curves are shown later in the manuscript.

\clearpage

\section{DISCOVERIES} \label{section_discovery}

The HSD survey in this work identified 38 new variable discoveries. Fourteen of the new discoveries are HSD binaries, including the compact binaries EVR-CB-001 and EVR-CB-004 both showing strong light curve variation due to ellipsoidal deformation effects from an unseen companion, and HW Virs EVR-CB-002 and EVR-CB-003 discussed below. We also detected 3 planet transit candidates, later shown to be false positives, appearing as potential planets because of a nearby source blended in the Evryscope pixel or due to a challenging, low airmass observational field. We found several reflection effect HSD binaries, and other spectroscopically confirmed HSD discoveries that exhibit sinusoidal like variability. The survey also revealed several other potentially high-priority targets for followup, which we discuss in \S~\ref{section_other_discovery_comments}

\subsection{Compact Binaries}

\begin{figure}[h]
\includegraphics[width=1.0\columnwidth]{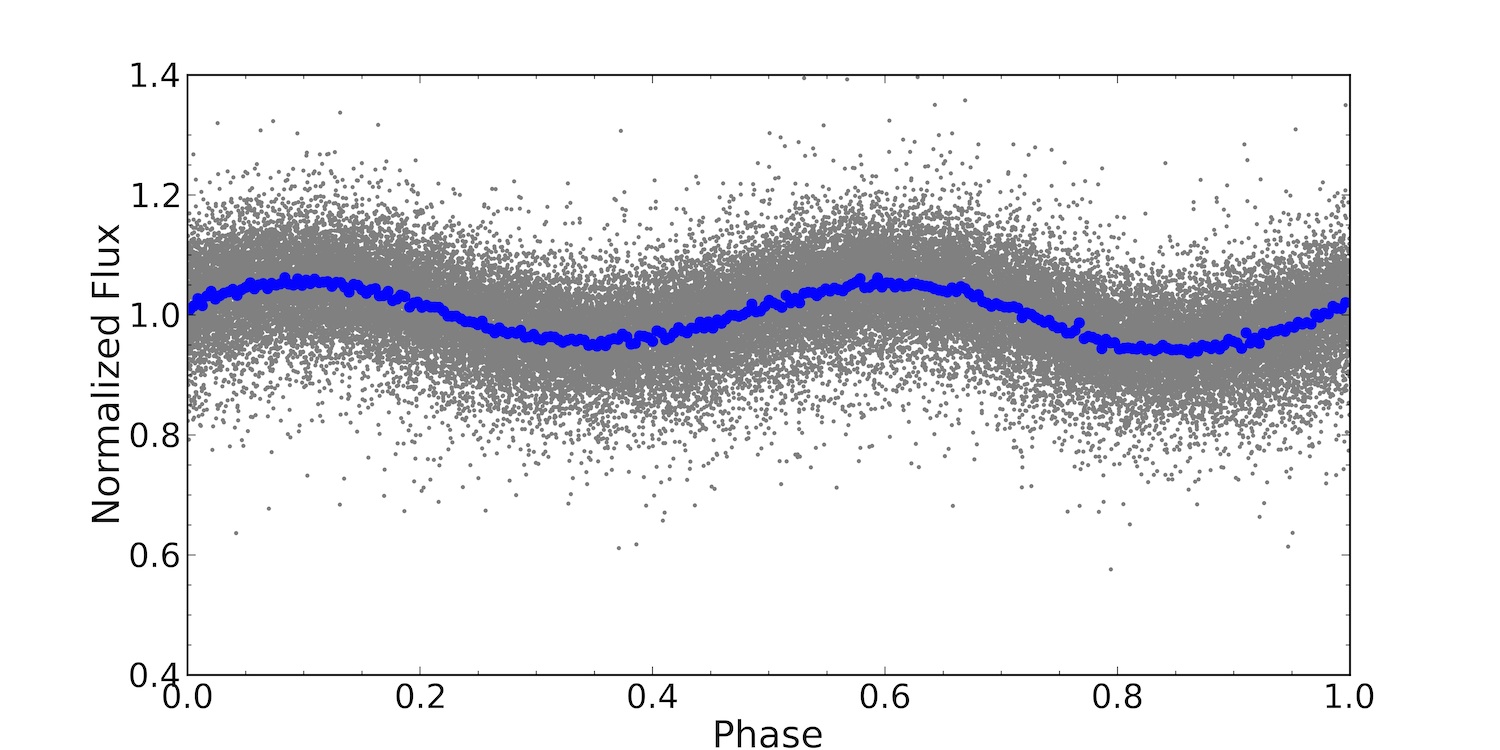}
\includegraphics[width=1.0\columnwidth]{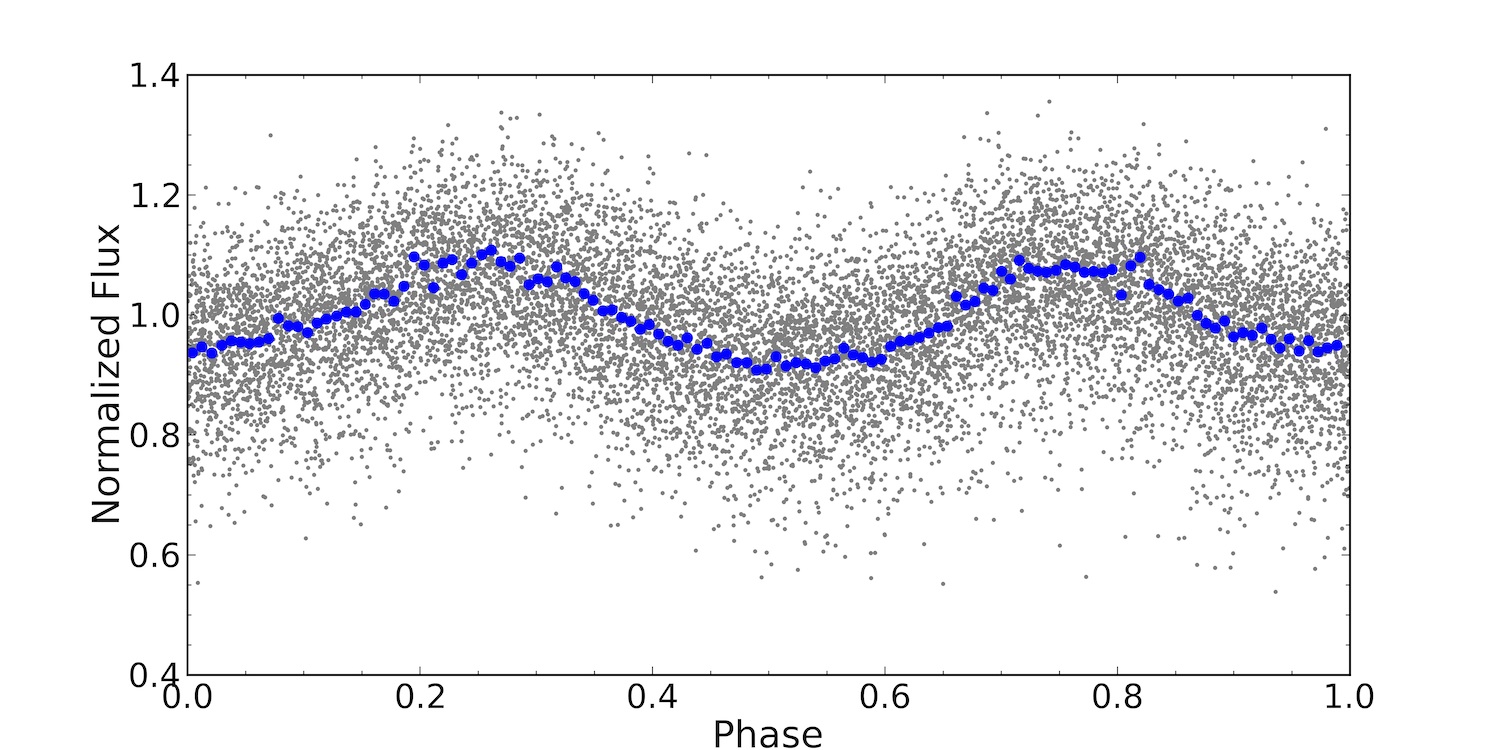}
\caption{\textit{Top:} The Evryscope light curve of EVR-CB-001 a 2.34 hour compact binary, with a very low mass unseen WD companion and a pre-He WD primary. \textit{Bottom:} The Evryscope light curve of EVR-CB-004 a 6.08 hour compact binary. Grey points = 2 minute cadence, blue points = binned in phase. The systems show ellipsoidal deformation of the primaries due to the unseen companions, as well as Doppler boosting and gravitational limb darkening.}
\label{fig:compact_binaries}
\end{figure}

EVR-CB-001 \citep{2019ApJ...883...51R} and EVR-CB-004 (Ratzloff et al., in prep), shown in Figure \ref{fig:compact_binaries}, are compact binary discoveries from the HSD survey, published in separate discovery papers with detailed followup and solutions. Both of these systems have a HSD spectral type primary and an unseen, degenerate companion. The variability in their light curves is dominated by the ellipsoidal deformation of the primary from the unseen companion, but also shows smaller amplitude effects due to Doppler boosting and gravitational limb darkening. We also recovered the only known system (CD-30 11223 \citealt{2041-8205-759-1-L25}) in our magnitude range and FoV. A summary of the compact binary discoveries is shown in Table \ref{tab:compact_bin}; we discuss the rarity of the systems in \S~\ref{section_rates}.

\begin{table}[h]
\caption{Compact Binaries}
\begin{tabular}{ l c c c c}
ID & RA & Dec & mag [G] & Period [h]\\
\hline
New Discoveries\\
EVR-CB-001$^a$ & 132.0648 & -74.3152 & 12.58 & 2.3425\\
EVR-CB-004$^b$ & 133.3023 & -28.7684 & 13.13 & 6.0842\\
\hline
Known Recoveries\\
CD-30 11223$^c$ & 212.8173 & -30.8844 & 12.32 & 1.1755\\
\hline
\multicolumn{5}{l}{$^a$\cite{2019ApJ...883...51R}, $^b$Ratzloff et al., in prep,}\\
\multicolumn{5}{l}{$^c$\cite{2041-8205-759-1-L25}}\\
\end{tabular}
\label{tab:compact_bin}
\end{table}

\subsection{HW Vir systems} \label{section_hwvir_discovery}

\begin{figure}[h]
\includegraphics[width=1.0\columnwidth]{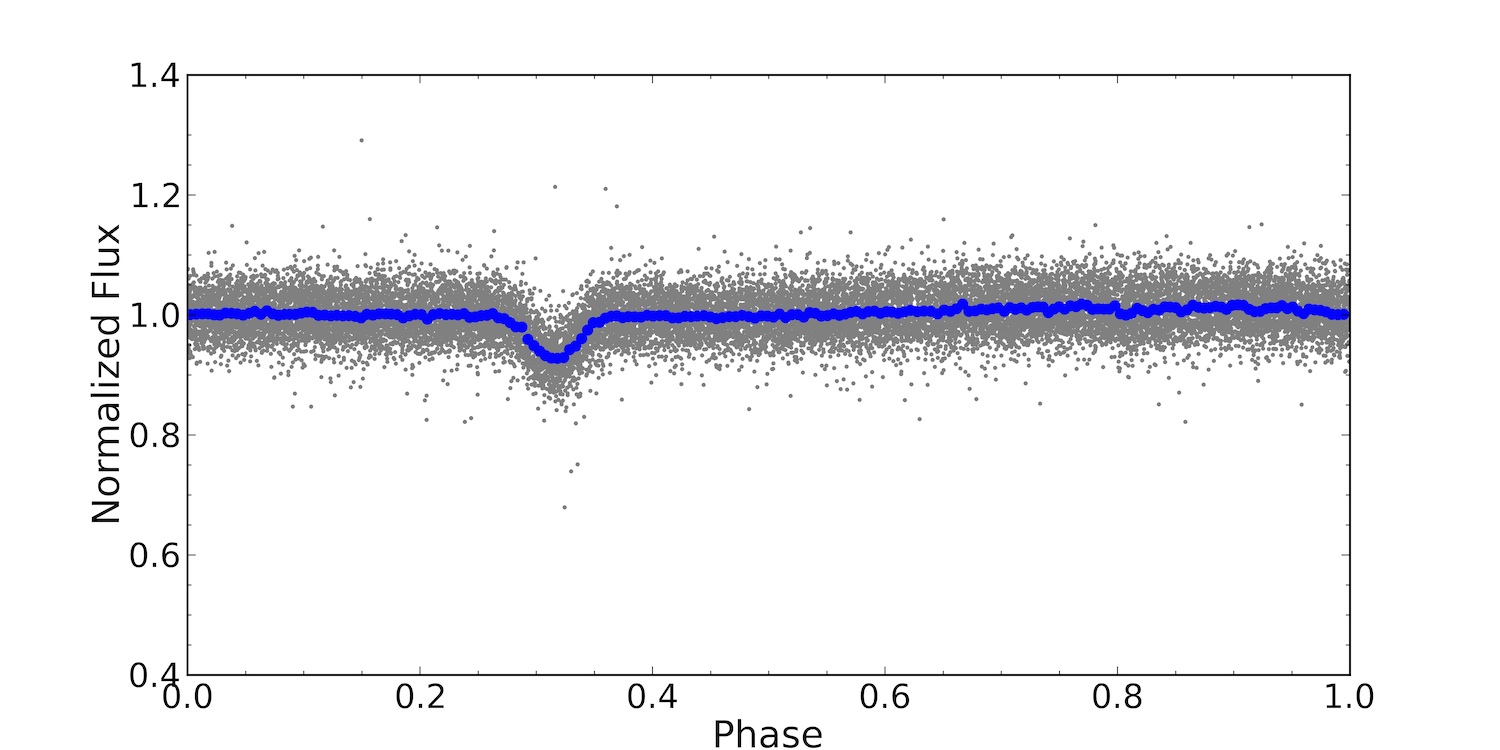}
\includegraphics[width=1.0\columnwidth]{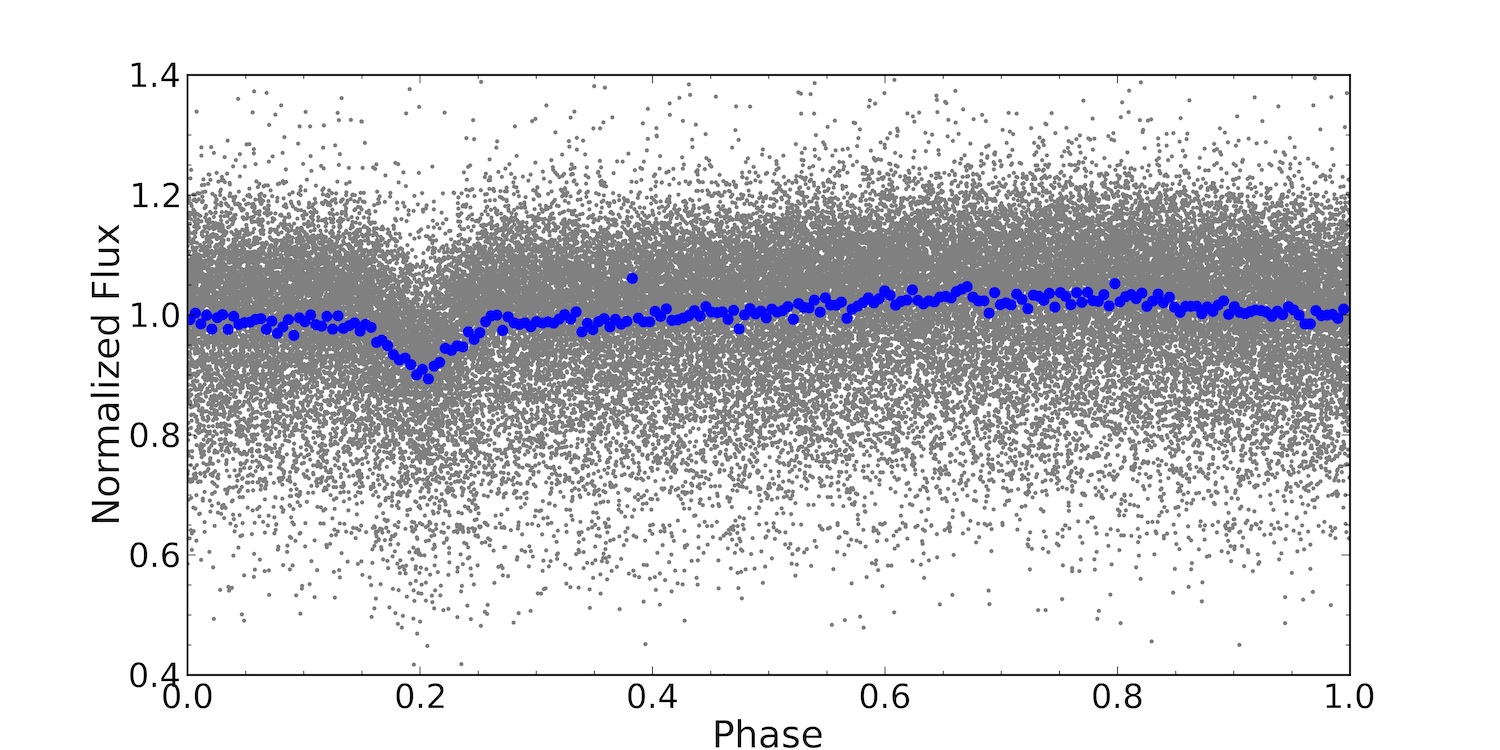}
\caption{\textit{Top:} The Evryscope light curve of EVR-CB-002 a 6.59 hour HW Vir. \textit{Bottom:} The Evryscope light curve of EVR-CB-003 a 3.16 hour HW Vir. Grey points = 2 minute cadence, blue points = binned in phase. The systems were challenging discoveries due to blended sources or crowded fields with high-airmass observations. Followup with higher resolution instruments separated the sources and revealed the HW Vir signals (Ratzloff et al., in prep).}
\label{fig:hwvirs_eslc}
\end{figure}

EVR-CB-002 and EVR-CB-003 (Ratzloff et al., in prep) are HW Vir discoveries from the HSD survey, with detailed followup and solutions published in a separate discovery paper. The discoveries are bright, southern sky systems facilitating followup and precise solutions. EVR-CB-002 features a high mass secondary (for HW Vir systems) of $M_{2}>\approx0.2M_{\odot}$ and EVR-CB-003 shows a very high reflectivity for HW Vir systems. We also recover all 5 of the known systems in our magnitude range and FoV (see \cite{2018A&A...614A..77S} for a list of the 20 known, solved HW Vir systems). The Evryscope discovery light curves are shown in Figure \ref{fig:hwvirs_eslc} and the recovery of HW Vir (the namesake system) is shown in Figure \ref{fig:det_outlier}. A summary of the HW Vir discoveries is shown in Table \ref{tab:hw_virs}; we estimate the occurrence rate of the systems in \S~\ref{section_rates}.

\begin{table}[h]
\caption{HW Vir Systems}
\begin{tabular}{ l c c c c}
ID & RA & Dec & mag [G] & Period [h]\\
\hline
New Discoveries\\
EVR-CB-002$^a$ & 79.9486 & -19.2816 & 13.61 & 6.5901\\
EVR-CB-003$^{a,b}$ & 210.4810 & -75.2260 & 13.53 & 3.1567\\
\hline
Known Recoveries\\
HW Vir$^c$ & 191.0843 & -8.6713 & 10.61 & 2.8013\\
AADor$^d$ & 82.9182 & -69.8839 & 11.16 & 6.2769\\
NSVS 14256825$^e$ & 305.0019 & 4.6324 & 13.25 & 2.6490\\
NYVir$^f$ & 204.7006 & -2.0303 & 13.39 & 2.4244\\
EC10246-2707$^g$ & 156.7353 & -27.3825 & 14.44 & 2.8443\\
\hline
\multicolumn{5}{l}{$^a$Ratzloff et al., in prep,}\\
\multicolumn{5}{l}{$^b$also identified in Jayasinghe et al. (in prep) as a general variable}\\
\multicolumn{5}{l}{(ASASSN-V J140155.45-751333.7), $^c$\cite{1986IAUS..118..305M},}\\
\multicolumn{5}{l}{$^d$\cite{1978Obs....98..207K}, $^e$\cite{2007IBVS.5800....1W}, $^f$\cite{1998MNRAS.296..329K},}\\
\multicolumn{5}{l}{$^g$\cite{2013MNRAS.430...22B}}\\
\end{tabular}
\label{tab:hw_virs}
\end{table}

\subsection{Planet Transit Candidates}\label{section_planet_transit_candidates}

\begin{figure}[h]
\includegraphics[width=1.0\columnwidth]{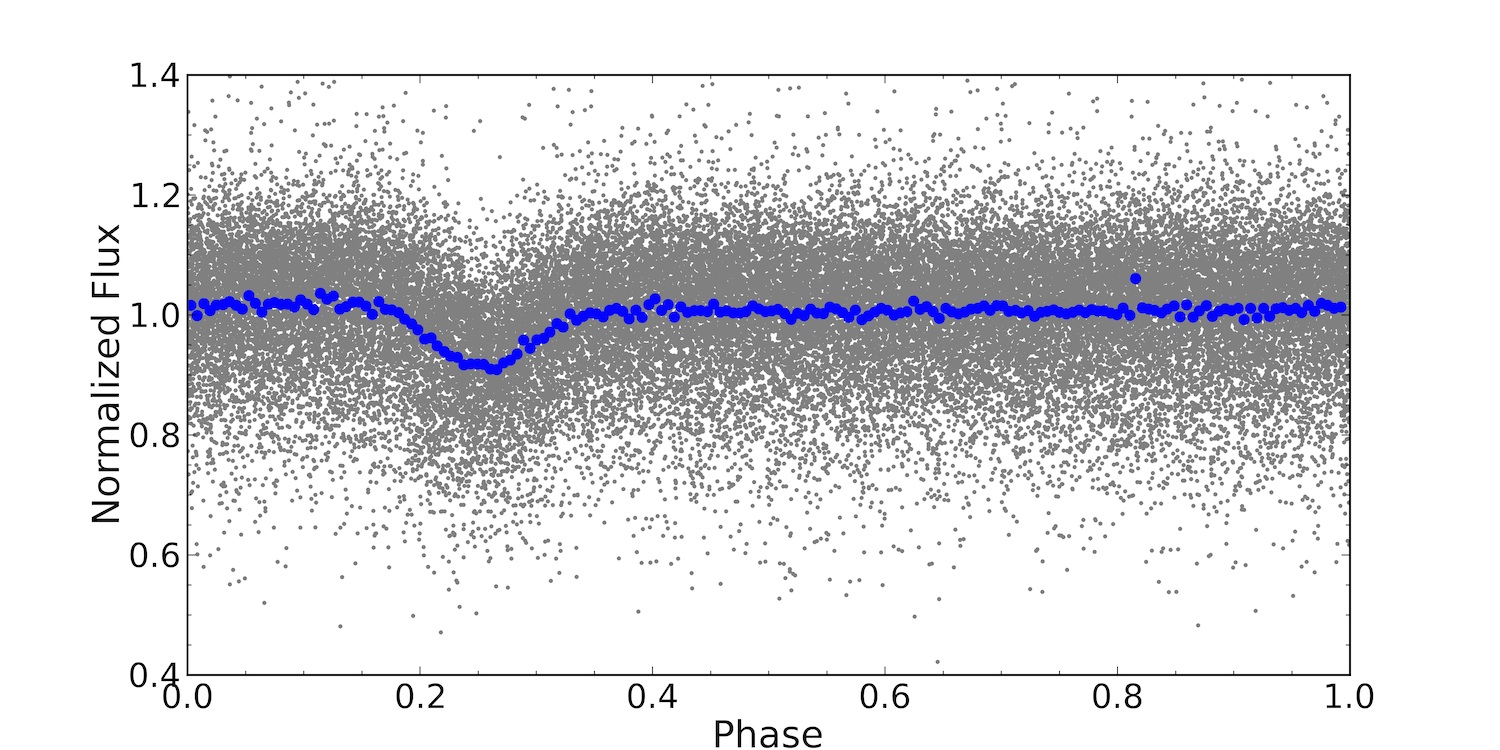}
\caption{The Evryscope light curve of a 2.68 hour transiting system, originally flagged as a HSD planet candidate. Grey points = 2 minute cadence, blue points = binned in phase. Followup revealed the target to instead be a suspected Cataclysmic Variable. We discovered two other planet candidates in the HSD search, which were later shown to be stellar in nature. These recoveries demonstrate the ability of our HSD survey to reach transit signals of sub-Jupiter size planets, from light curves with similar astrophysical signals.}
\label{fig:planet_candidates}
\end{figure}

Three planet candidates were identified from the HSD search, all showing transit times of $\approx$ 20 minutes and depths of less than 10\%. The discovery light curves do not show signs of secondary eclipses or grazing transits. Assuming the host star for each system is a HSD with a 0.2 $R_{\odot}$ radius, the transiting object would be sub-Jupiter in size. Photometric followup revealed two of the candidates to be the HW Vir systems (EVR-CB-002 and EVR-CB-003) presented in \S~\ref{section_hwvir_discovery}. The actual transit depths are much deeper than in the Evryscope discovery light curves because of a nearby star that was blended in the Evryscope pixels (EVR-CB-002) and a high-airmass observing field (EVR-CB-003). Spectroscopic followup revealed the final candidate to be a suspected Cataclysmic Variable, but with odd HSD like features in the spectrum. Neither the discovery or the followup light curves show signs of outbursts. The Evryscope light curve is shown in Figure \ref{fig:planet_candidates}. We are still exploring the nature of this candidate.

As there are no known exoplanets transiting HSDs, we are forced to rely completely on simulations to test our recovery algorithms and to estimate our detection efficiency. The transit signals of these three candidates are very similar to expected HSD transiting gas giant planets (slightly smaller than Jupiter size), and demonstrate the ability of our HSD survey to recover fast transit planet signals in actual Evryscope light curves.

\subsection{Reflection Effect or Partially Eclipsing Binaries} \label{section_rb_ec015781743}

\begin{figure*}[htb]
\includegraphics[width=2.0\columnwidth]{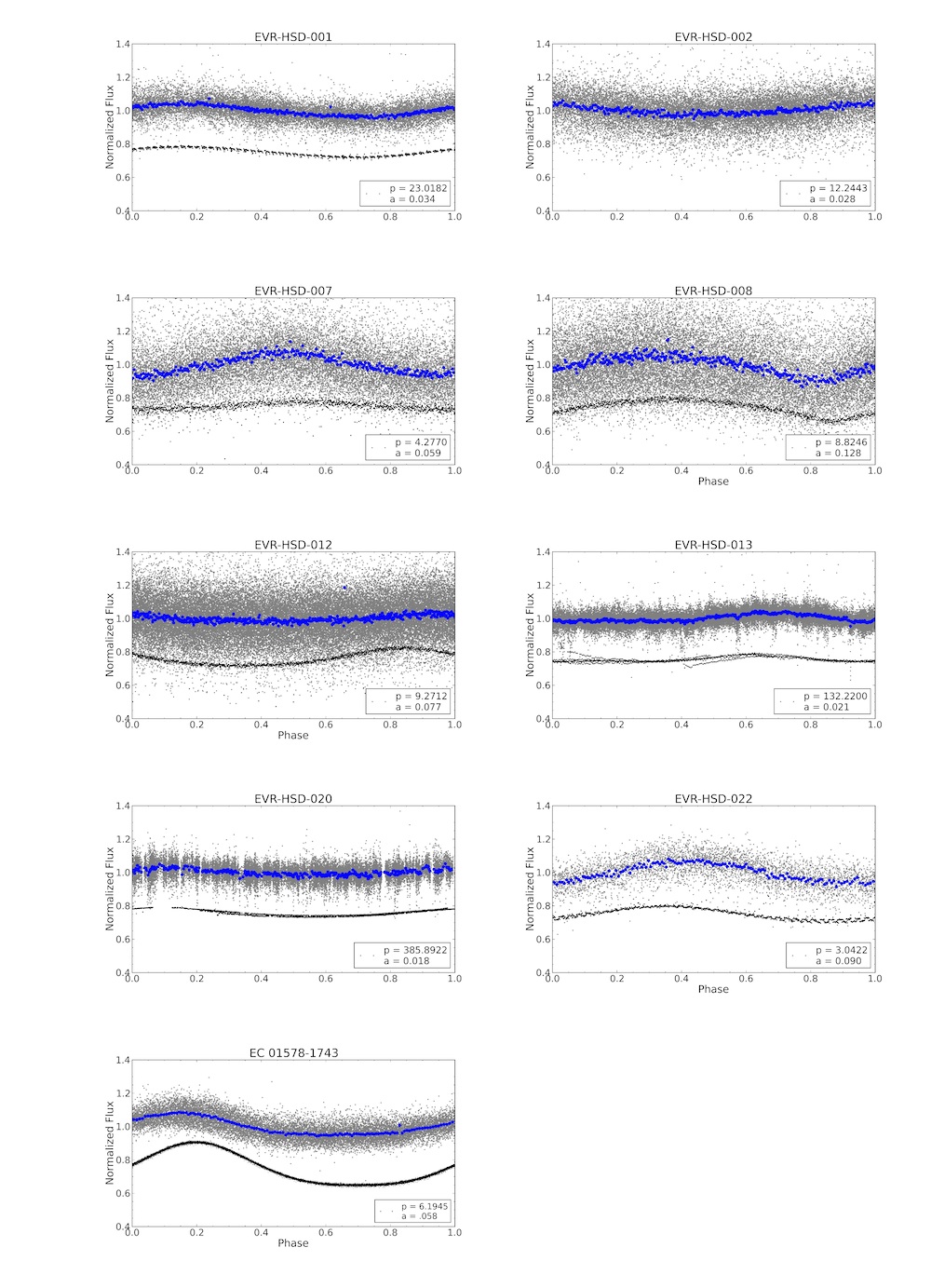}
\caption{The Evryscope light curves of HSD variable discoveries showing reflection or sinusoidal signals with periods ranging from 3 to 386 hours. The period and amplitudes (fractional change from mean to peak measured in the Evryscope light curves) shown are from the best LS fit. Grey points = 2 minute cadence, blue points = binned in phase. The TESS light curves (where available and indicated with the black points) are shown for comparison with a .25 offset in normalized flux for better visualization.}
\label{fig:var_disc_1}
\end{figure*}

We discovered the HSD reflection binary EC 01578-1743, first presented in \cite{2019AAS...23346403W}, and reported in detail here. We discovered 9 additional HSD variables with periods ranging from 3 to 386 hours. The photometric variation is likely due to binary effects. A summary of the results is shown in Table \ref{tab:ref_bin}. The Evryscope light curves are shown in Figure \ref{fig:var_disc_1}. The discovery amplitudes and periods are from the best LS detection and fit to the Evryscope light curves. The LS detection powers are significantly above the survey average (32.2 compared to 15). The TESS light curves are shown for comparison wherever available. Additional discovery details including spectral types from the best fits to the spectra are shown in Table \ref{tab:var_discoveries_sum}, along with a listing of all discoveries from this work.

The HSD survey recovered 4 of the 6 known short period HSD reflection binaries in our magnitude range and FoV. The two known systems that were not recovered, CPD-64481 and PHL 457 \citep{2014A&A...570A..70S}, were missed due to low amplitude (sub-1\%) variability and a close source that was blended in the Evryscope pixels. A list of known, solved HSD binaries showing reflection effects can be found in \cite{2015A&A...576A..44K}. The HSD search also recovered four known eclipsing binaries: V1379Aql \citep{1982IBVS.2215....1H} a HSD/K-giant, EC21049-5649 \citep{2009ApJ...696..870D}, EC23257-5443 \citep{10.1093/mnras/stw916} and GALEX J175340.5-500741 \citep{10.1093/mnras/stv821} HSD/Fs. A list of known HSD eclipsing binaries can be found in \cite{10.1093/mnras/stv821}.

\begin{table}[h]
\caption{HSD Reflection Effect or Eclipsing Binaries}
\begin{tabular}{ l c c c c}
ID & RA & Dec & mag [G] & Period [h]\\
\hline
New Discoveries\\
EC 01578-1743$^{a,b}$ & 30.0553 & -17.4788 & 12.05 & 6.1945\\
EVR-HSD-001 & 40.2665 & -19.0032 & 12.55 & 23.0182\\
EVR-HSD-002 & 97.1064 & -18.7484 & 13.19 & 12.2443\\
EVR-HSD-007 & 271.7181 & -43.5589 & 13.47 & 4.2769\\
EVR-HSD-008 & 73.7044 & -65.8895 & 14.87 & 8.8246\\
EVR-HSD-012 & 151.4384 & -63.5280 & 13.09 & 9.2712\\
EVR-HSD-013 & 158.7382 & -53.8975 & 11.58 & 132.223\\
EVR-HSD-020 & 295.0117 & -49.4531 & 12.03 & 385.89\\
EVR-HSD-022 & 133.3023 & -28.7684 & 13.13 & 3.0422\\
\hline
Known Recoveries\\
TYC 7709-376-1$^c$ & 155.8412 & -37.6166 & 11.71 & 3.3425\\
TW Crv$^d$ & 180.0235 & -19.0344 & 15.03 & 7.8629\\
KV Vel$^e$ & 163.6690 & -48.7841 & 12.18 & 8.5709\\
BPS CS22169-0001$^f$ & 59.0972 & -15.1554 & 12.86 & 5.2057\\
J175340.5-500741$^g$ & 268.4189	& -50.1284 & 12.88 & 2.1778\\
V1379Aql$^h$ & 294.9117 & -6.0637 & 7.81 & 624.77\\
EC21049-5649$^i$ & 317.1796 & -56.6181 & 14.37 & 6.3976\\
EC23257-5443$^j$ & 352.1339 & -54.4532 & 14.53 & 6.6334\\
\hline
\multicolumn{5}{l}{$^a$\cite{2019AAS...23346403W}, $^b$(also noted as an unidentifiable variable}\\
\multicolumn{5}{l}{ASAS J020013-1728.7) \cite{2002AcA....52..397P}, }\\
\multicolumn{5}{l}{$^c$(also known as ASAS 102322-3737.0) \cite{2013AA...553A..18S}, }\\
\multicolumn{5}{l}{$^d$(also known as EC11575-1845) \cite{1995MNRAS.275..100C}, }\\
\multicolumn{5}{l}{$^e$\cite{1986AJ.....91.1372L}, $^f$\cite{2005AA...442.1023E}, }\\
\multicolumn{5}{l}{$^g$\cite{10.1093/mnras/stv821}, $^h$(also known as HD 185510) \cite{1982IBVS.2215....1H}, }\\
\multicolumn{5}{l}{$^i$(also known as DDE 98) \cite{2009ApJ...696..870D}, $^j$\cite{10.1093/mnras/stw916}}\\

\end{tabular}
\label{tab:ref_bin}
\end{table}

\clearpage

\subsection{Highlighted Discoveries}\label{section_other_discovery_comments}

\subsubsection{EC 01578-1743}

From the best fit to the SOAR ID spectra using \textit{Astroserver} \citep{2017OAst...26..179N}, we measure $T_{\rm eff} = 31,980 K$, $\log{g}=5.78$ cm\,s$^{-2}$. We classify EC 01578-1743 as an sdB, and identify it as a reflection effect HSD binary. Initial light curve and radial velocity solutions indicate a late M-dwarf companion. A full, detailed solution of EC 01578-1743 including precise fitting of the TESS and Evryscope light curves will be presented in an upcoming work (Schaffenroth, et al., in prep).

\subsubsection{EVR-HSD-001, EVR-HSD-002, EVR-HSD-007, EVR-HSD-022}

The variables identified here are short period, moderate amplitudes, and with binary reflection effect signals. Each has been spectroscopically confirmed as an sdB. The bright magnitudes will also aid in photometric or radial velocity followup. 

\subsubsection{EVR-HSD-008}

From the best spectral fit, EVR-HSD-008 is a hot sdB or post GB star but with a very high projected rotational velocity (which could also be indicative of other broadening mechanisms such as magnetic, instrumentational, or orbital smearing that might be seen in a higher resolution spectrum). The short period and very distinct features are confirmed in the TESS light curve. EVR-HSD-008 is also identified as a potential post AGB candidate in \cite{2015MNRAS.454.1468K}. There is also a slight phase offset between the Evryscope and TESS light curves that we are unable to explain, and requires further followup. This target is a strong candidate for RV measurements and additional analysis (Galliher et al., in prep).

\subsubsection{EVR-HSD-012}

A very strong sdB reflection candidate, with the 9.2712 hour period confirmed with the TESS light curve. The TESS amplitude is higher than that seen in the Evryscope light curve, potentially a consequence of the reflection effect observed in different filters.

\subsubsection{EVR-HSD-013}

An sdB reflection binary candidate with a long 5.5 day period. From the spectral fit, this is \textit{potentially} a double line system which would offer a rare opportunity to measure the mass of the HSD directly. Followup with a higher resolution spectrum is needed to confirm or reject this hypothesis.

\subsubsection{EVR-HSD-020}

A difficult to find long period (385.8922 hour) variable with a reflection like shape. We note here that since EVR-HSD-020 is quite a long period variable, the TESS light curve is from a single sector (13 the only one available at the time of our survey). The reflection shape and longer period could be indicative of an earlier main sequence companion. The spectral fit classifies this star as an sdB or BHB. 

\subsection{Spectroscopically Confirmed HB and B Variables}

\begin{figure*}[htb]
\includegraphics[width=2.0\columnwidth]{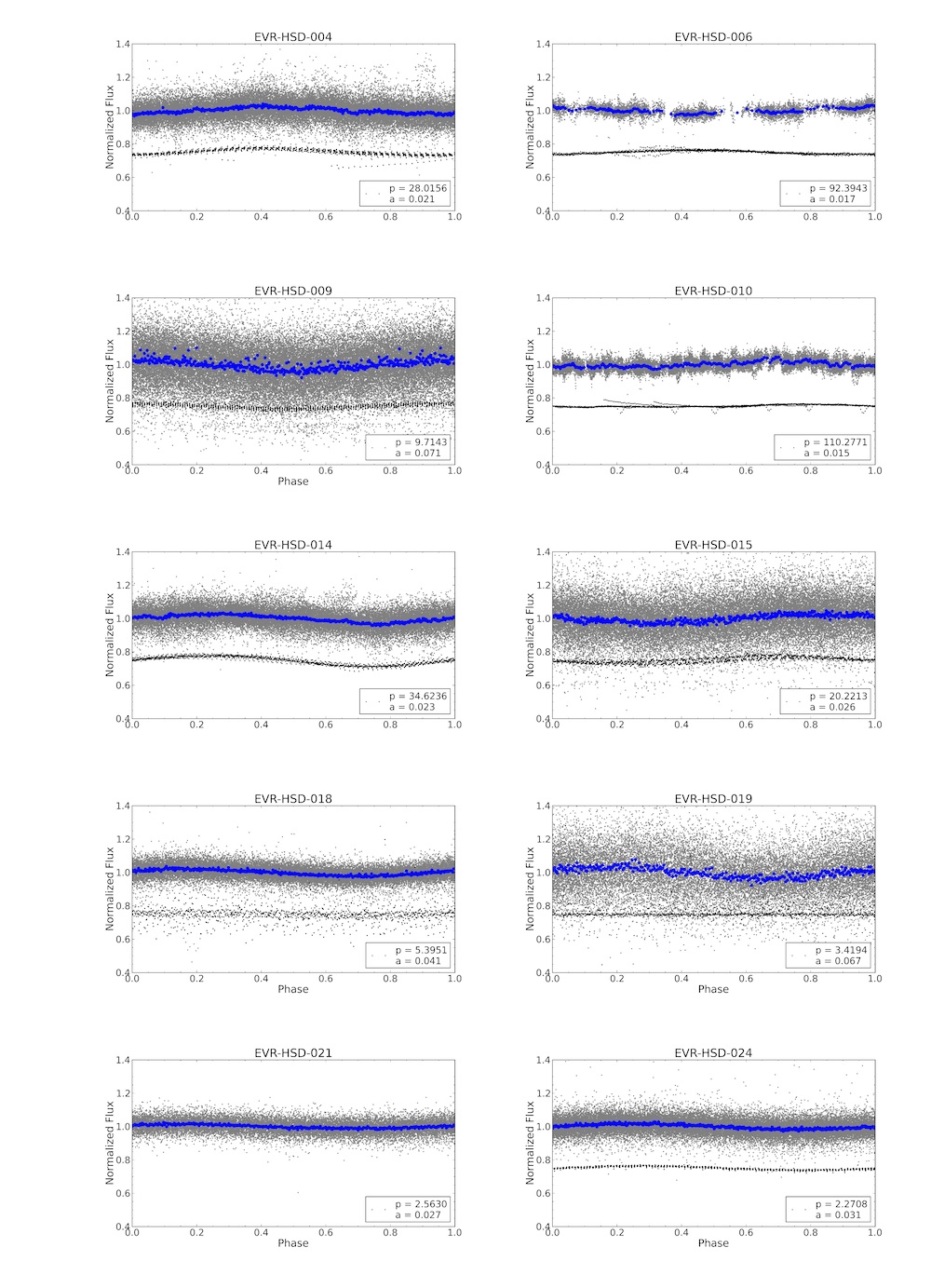}
\caption{The Evryscope light curves of variable discoveries showing reflection or sinusoidal signals with periods ranging from 2.5 hours to 110 hours. The period and amplitudes (fractional change from mean to peak measured in the Evryscope light curves) shown are from the best LS fit. Grey points = 2 minute cadence, blue points = binned in phase. The TESS light curve (where available and indicated with the black points) are shown for comparison with a .25 offset in normalized flux for better visualization.}
\label{fig:var_disc_2}
\end{figure*}

The remaining spectroscopically confirmed variable discoveries are horizontal branch (HB) and B stars. The light curves show sinusoidal or reflection features in periods ranging from a few hours to nearly 5 days. The results are shown in Table \ref{tab:b_stars} and the light curves in Figure \ref{fig:var_disc_2}. Additional discovery details including spectral types from the best fits to the spectra are shown in Table \ref{tab:var_discoveries_sum}, along with a listing of all discoveries from this work.

\begin{table}[h]
\caption{HB and B Variables}
\begin{tabular}{ l c c c c}
ID & RA & Dec & mag [G] & Period [h]\\
\hline
New Discoveries\\
EVR-HSD-004 & 194.6749 & -35.3798 & 12.22 & 28.0156\\
EVR-HSD-006$^a$ & 263.4460 & -70.9357 & 10.60  & 92.39\\
EVR-HSD-009 & 75.0706 & -65.8073 & 15.09 & 9.7143\\
EVR-HSD-010 & 107.7860 & -7.1754 & 10.22 & 110.277\\
EVR-HSD-014 & 162.6241 & -39.7614 & 12.00 & 34.6236\\
EVR-HSD-015 & 170.0517 & -57.3402 & 12.98 & 20.2213\\
EVR-HSD-018 & 267.0839 & -49.6796 & 11.81 & 5.3951\\
EVR-HSD-019 & 285.2468 & -35.4992 & 13.14 & 3.4194\\
EVR-HSD-021 & 301.6089 & -6.5474 & 11.41 & 2.5630\\
EVR-HSD-024 & 211.1865 & -49.2117 & 11.90 & 2.2708\\
\hline
\multicolumn{5}{l}{$^a$(also CPD-702387, noted as a potential OB star)}\\
\multicolumn{5}{l}{\cite{1995PASP..107..846D}}\\

\end{tabular}
\label{tab:b_stars}
\end{table}

\subsection{Cataclysmic and other Outbursting Variables}

Although not a focus of the surveys, we recovered several known Cataclysmic Variables and Novae. The nature of these systems (WD host and compact binaries) leads in some cases to light curve features similar to those expected from a HSD planet transit. The short periods, depths, and shapes (but with somewhat longer transits) are comparable to the HSD planet simulations; and demonstrates in a separate target group the ability of our detection algorithms to recover transit signals in actual Evryscope light curves.

\begin{table}[h]
\caption{Cataclysmic Variables and Novae }
\begin{tabular}{ l c c c c}
ID & RA & Dec & mag [G] & Period [h]\\
\hline
Known Recoveries\\
AO Psc$^a$ & 343.8249 & -3.1778 & 13.23 & 3.5910\\
UU Aqr$^b$ & 332.2740 & -3.7716 & 13.58 & 3.9259\\
TX Col$^c$ & 85.8340 & -41.0318 & 15.62 & 5.7192\\
EC21178-5417$^d$ & 320.3606 & -54.0763 & 13.74 & 3.7087\\
RR Pic$^e$ & 98.9003 & -62.6401 & 12.41 & 3.4806\\
SV CMi$^f$ & 112.7851 & 5.9802 & 16.03 & 3.7440\\
\hline
\multicolumn{5}{l}{$^a$\cite{1980MNRAS.193P..25G}, $^b$\cite{2001PASP..113..764D}, $^c$\cite{1986ApJ...311..275T}, }\\
\multicolumn{5}{l}{$^d$\cite{2001PASP..113..764D}, $^e$\cite{1939PA.....47..538M}, $^f$\cite{2001PASP..113..764D} }\\

\end{tabular}
\label{tab:cv_recoveries}
\end{table}


\subsection{Peculiar Discoveries}

\subsubsection{EVR-HSD-010}

\begin{figure}[h]
\includegraphics[width=1.0\columnwidth]{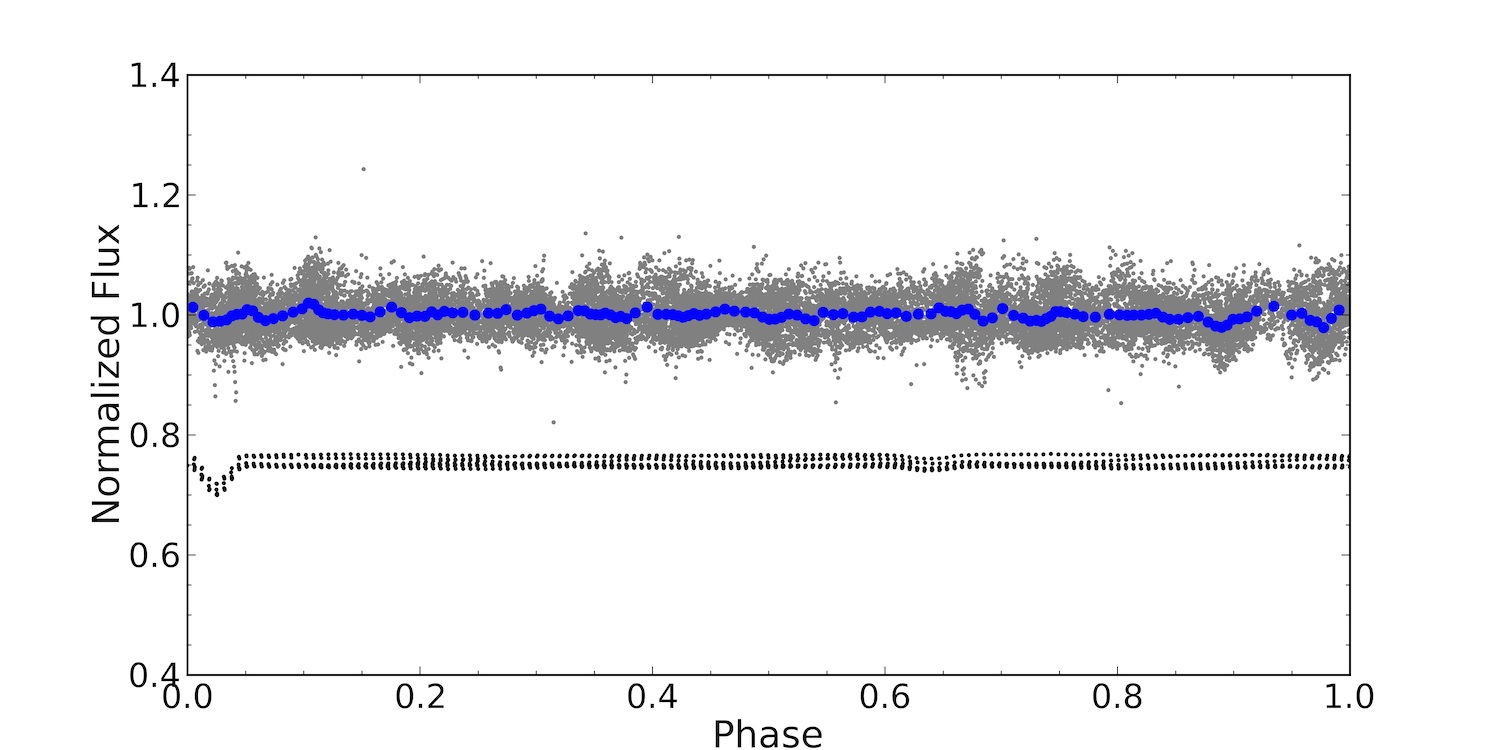}
\caption{The Evryscope and TESS light curves of the multi-variable system EVR-HSD-010, folded here on the 77.9885 hour eclipsing period. Grey points = 2 minute cadence, blue points = binned in phase. The TESS light curve is shown with a .25 offset in normalized flux for better visualization.}
\label{fig:H16764}
\end{figure}

EVR-HSD-010 is a HB star with the spectral fit indicating a lower temperature and surface gravity than a typical HSD. The 110.277 hour sinusoidal (or possible reflection) variability is confirmed in the TESS light curve but at a lower amplitude. Also visible in the TESS light curve are shallow (4\%) eclipses at a different period (77.9885 hours). The $\approx$ 4 hour duration shallow eclipse, HB star type, and period suggest a reasonably large (solar radius or larger) primary and small, dim secondary (most likely a late M-dwarf). The bright magnitude would aid in further followup of this system (Galliher et al., in prep). We show the light curve folded to the eclipse period in Figure \ref{fig:H16764}.

\subsubsection{EVR-HSD-019}

The EVR-HSD-019 Evryscope light curve indicates short period reflection effect or sinusoidal like variability, however the effect is not present in the TESS light curve. This could be due to a color effect, or it could be a systematic in the Evryscope light curve instead of an astrophysical signal.

\subsection{CPD-634369}

\begin{figure}[h]
\includegraphics[width=1.0\columnwidth]{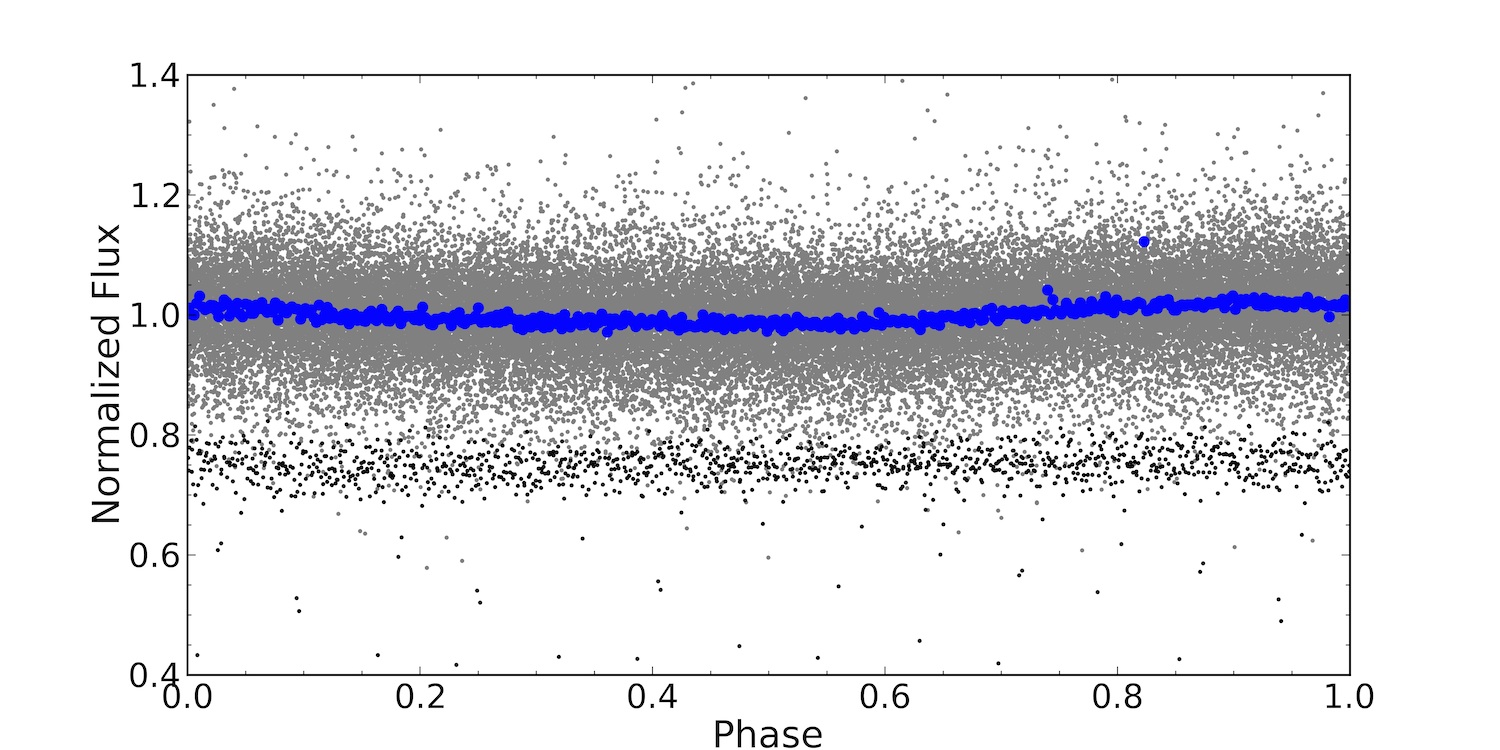}
\includegraphics[width=1.0\columnwidth]{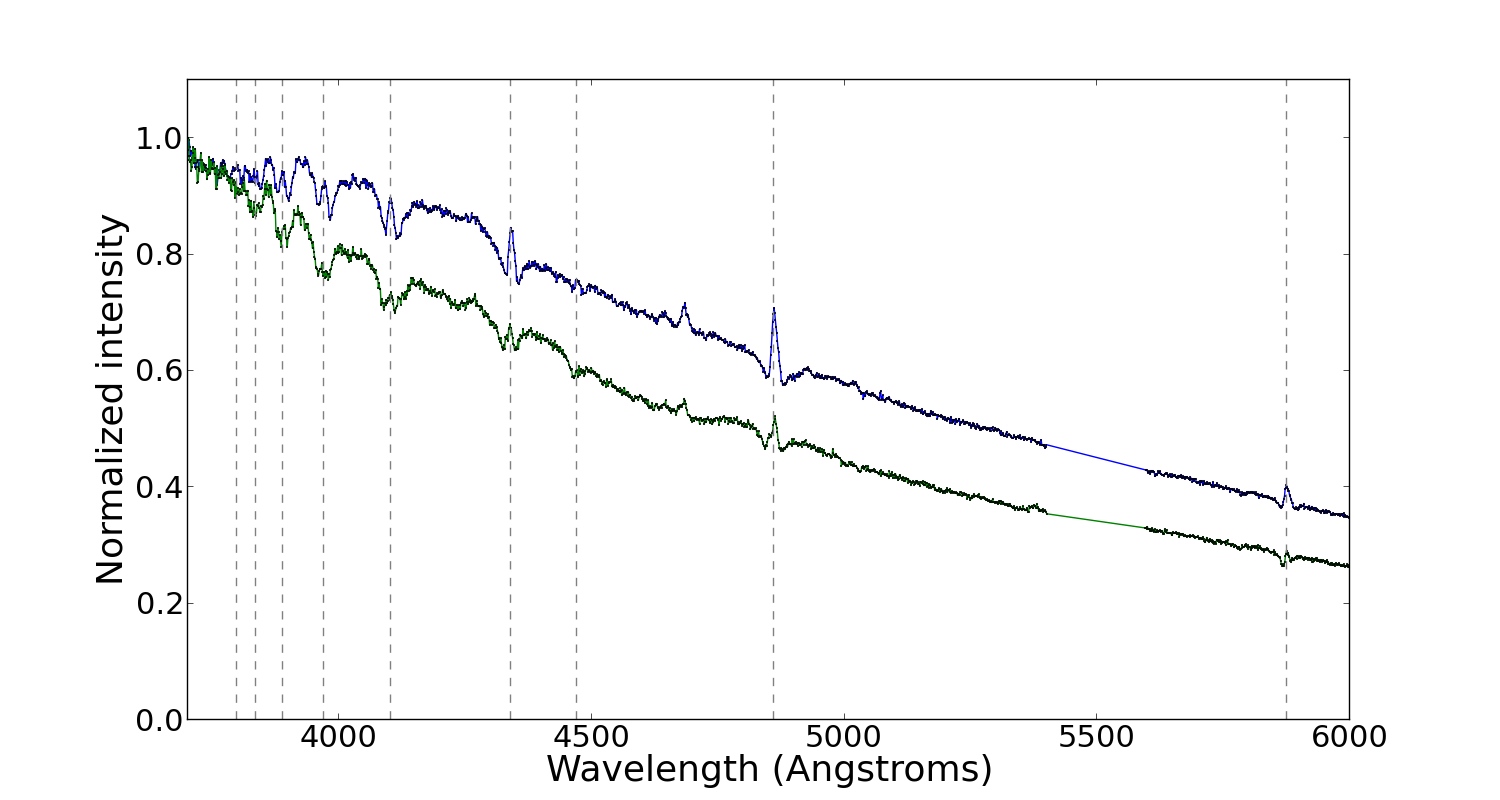}
\caption{\textit{Top:} The Evryscope light curve of the potential CV or debris disc CPD-634369, folded here on the 3.2177 hour period. Grey points = 2 minute cadence, blue points = binned in phase. The TESS light curve (black points) is shown with a .25 offset in normalized flux for better visualization. \textit{Bottom:} The SOAR ID spectra, showing broadened absorption features and a high temperature consistent with a WD but with emissions indicative of mass transfer. The emission features change in amplitude as seen by comparing spectra taken in March 2019 (Blue) and September 2019 (Green). H-$\alpha$ to H-10(dashed lines) are shown for reference.}
\label{fig:Debris_disc}
\end{figure}

CPD-634369 is a short period variable that shows peculiar spectral features, which warrant further investigation. Our initial followup SOAR spectra (with measurements taken over the photometric period), show broad absorption features and superimposed emissions that change over the period cycle. The photometric and spectral features are shown in Table \ref{tab:disc} and in Figure \ref{fig:Debris_disc}. We identify CPD-634369 as a potential cataclysmic variable (CV) in a low-mass transfer state, perhaps similar to V379 Vir or comparable systems \citep{2019arXiv190713152P, 2005ApJ...630L.173S, 2019MNRAS.484.2566L, 2011MNRAS.416.2768S}. The low amplitude emission lines suggest mass loss perhaps with an accretion disc, while it is also possible the object has a debris disc. A very hot, blue star CPD-634369 was noted as a OB candidate in \cite{1995PASP..107..846D}. We identify the object as a probable WD primary, cataclysmic-variable-like oscillations, and possibly with a debris disc. Additional spectroscopic and RV followup is necessary for confirmation (Galliher and Barlow et al., in prep).

\begin{table}[h]
\caption{CV candidate / WD Debris Disc}
\begin{tabular}{ l c c c c}
ID & RA & Dec & mag [G] & Period [h]\\
\hline
New Discoveries\\
CPD-634369 & 274.7516 & -63.3006& 12.30 & 3.2177\\
\hline
\end{tabular}
\label{tab:disc}
\end{table}

\subsection{Other Discoveries}

Other discoveries from the HSD survey are shown in the appendix. They are suspected misclassified stars (from only 1 source, see \S~\ref{section_list_perf}) - most likely A or B stars. While not the focus of the surveys, there is a variety of variability including reflection binaries, eclipsing binaries, sinusoidal variables, and peculiar variables. The best period and amplitude fits are also provided.

\subsection{Discoveries Summary} \label{section_discovery_var}

Discoveries from this section are summarized in Table \ref{tab:var_discoveries_sum}. GAIADR2 data is listed for the ID cross-reference, RA, Dec, and mag[G]. The effective temperature, surface gravity, and projected rotational velocity are determined from the best fit to the SOAR ID spectra (see \S~\ref{section_followup}). The projected rotation velocity is difficult to estimate from low resolution spectra. Where we could not determine a value, we list only an upper limit. However, there are cases where the fit procedure converged on very high values. These we interpret as the residual effect of another unresolved broadening processes, such as smearing due to orbital movement in close binaries. We use the values determined from the spectral fits to determine a spectral type and consider the light curve variation as a reasonableness check. Periods listed are from the best BLS or LS fit from the light curve discovery. 

\LongTables
\begin{sidewaystable*}
\caption{Discovery Table \\ Summary of Discoveries from this work (Spectroscopically Confirmed Spectral Type)}
\centering
\begin{tabular}{ l c c c c c c c c c c}
ID & ID (GaiaDR2+) & RA & Dec & mag [G] & $T_{\rm eff}$ [K] & $\log{g}$ & $\log{n{\rm He}/n{\rm H}}$ & \textit{v}sin\textit{i} [km/s] & SpT & Period [h]\\
(ES Internal ID) & or common name\\
\hline
\bf{HSD Spectral Type}\\
\bf{Compact Binaries}\\
EVR-CB-001 & 5216785445160303744 & 132.0645 & -74.3151 & 12.581 $\pm.003$ & 18,500 $\pm500$ & 4.96 $\pm.04$ & -1.43 $\pm.03$ & 112 $\pm4$ & pre-He WD & 2.3425217(5)\\
(H06771)\\
EVR-CB-004 & 5642627428172192640 & 133.3023 & -28.7683 & 13.127 $\pm.002$ & 41,800 $\pm1400$ & 4.63 $\pm.11$ & -- & -- & sdO? & 6.0842 $\pm.0001$\\
(H03930)\\
\bf{HW Virs}\\
EVR-CB-002 & 2969438206889996160 & 79.9486 & -19.2816 & 13.608 $\pm.003$ & 27,963 $\pm224$ & 5.39 $\pm.033$ & -- & -- & sdB & 6.590132(8)\\
(H03281)\\
EVR-CB-003 & 5790285036556643072 & 210.4805 & -75.2258 & 13.534 $\pm.009$ & 32,552 $\pm152$ & 5.78 $\pm.032$ & -- & -- & sdB & 3.1567 $\pm.0001$\\
(H00250)\\
\multicolumn{10}{l}{\bf{Reflection Effect and Other Variables}}\\
EC 01578-1743 & 5141474254479109376 & 30.0553 & -17.4788 & 12.052 $\pm.003$ & 31,980 $\pm100$ & 5.78 $\pm.02$ & -2.09 $\pm.04$ & 58 $\pm30$ & sdB & 6.19449 $\pm.00001$\\
(H00061) \\
EVR-HSD-001 & PHL1434 & 40.2665 & -19.0032 & 12.547 $\pm.003$ & 20,950 $\pm150$ & 4.83 $\pm.04$ & -2.87 $\pm.07$ & 52 $\pm20$ & sdB & 23.0182 $\pm.0002$\\
(H00098) \\
EVR-HSD-002 & 2940254694388783104 & 97.1064 & -18.7484 & 13.193 $\pm.002$ & 24,920 $\pm200$ & 5.30 $\pm.03$ & -2.98 $\pm.06$ & 198 $\pm90$ & sdB & 12.24434 $\pm.00005$\\
(H00182) \\
EVR-HSD-007 & 6724092123091015552 & 271.7181 & -43.5589 & 13.468 $\pm.006$ & 27,390 $\pm230$ & 5.40 $\pm.04$ & -2.04 $\pm.04$ & 140 $\pm15$ & sdB & 4.27691 $\pm.00001$\\
(H05026) \\
EVR-HSD-008 & 4662207272056081152 & 73.7044 & -65.8895 & 14.866 $\pm.004$ & 33,130 $\pm800$ & 4.41 $\pm.11$ & -1.10 $\pm.08$ & 528 $\pm23$ & pGB/sdB? & 8.82458 $\pm.00001$\\
(H17205) \\
EVR-HSD-012 & 5252661788043512320 & 151.4384 & -63.5280 & 13.087 $\pm.002$ & 29,060 $\pm260$ & 5.49 $\pm.05$ & -2.42 $\pm.03$ & $<20$ & sdB & 9.27121 $\pm.00002$ \\
(H06317) \\
EVR-HSD-013 & 5354100321320289024 & 158.7382 & -53.8975 & 11.578 $\pm.002$ & 14,660 $\pm100$ & 4.05 $\pm.08$ & -2.35 $\pm.23$ & $<50$ & sdB+X? & 132.223 $\pm.006$\\
(H05722) \\
EVR-HSD-020 & 6647178772143579392 & 295.0117 & -49.4531 & 12.033 $\pm.001$ & 23,620 $\pm150$ & 4.86 $\pm.03$ & -3.85 $\pm.34$ & 174 $\pm20$ & sdB/BHB? & 385.89 $\pm.06$ \\
(H05487) \\
EVR-HSD-022 & 5642627428172192640 & 133.3023 & -28.7684 & 13.127 $\pm.002$ & 40,160 $\pm330$ & 4.63 $\pm.06$ & -1.08 $\pm.04$ & $<40$ & sdO/? & 3.04223 $\pm.00002$\\
(H08435) \\
\bf{CV or Debris Disc}\\
(H06304) & CPD-634369 & 274.7516 & -63.3006 & 12.296 $\pm.003$ & $\approx$40,000 & $\approx$8.0 & $\approx$-2.3 & -- & HSD/WD+DD? & 3.21766 $\pm.00003$\\
\multicolumn{10}{l}{\bf{Non-variables in the ES LCs}}\\
(H00526) & EC10578-3116 & 165.0660 & -31.5462 & 14.44 & 34,300 $\pm130$ & 5.85 $\pm.03$ & -1.57 $\pm.03$ & 118 $\pm25$ & sdB & --\\
(H00056) & PG1352-023 & 208.7694 & -2.5061 & 12.06 & 43,140 $\pm1250$ & 5.97 $\pm.09$ & -1.87 $\pm.06$ & $<40$ & sdO & --\\

\hline
\bf{HB or B stars}\\
\bf{Variables}\\
EVR-HSD-004 & 6155138767433459840 & 194.6749 & -35.3798 & 12.217 $\pm.001$  & 18,250 $\pm130$ & 3.95 $\pm.03$ & -1.00 $\pm.03$ & 165 $\pm10$ & B3V & 28.0156 $\pm.0002$\\
(H04413) \\
EVR-HSD-006 & CPD-702387 & 263.4460 & -70.9357 & 10.596 $\pm.001$ & 27,190 $\pm890$ & 3.89 $\pm.08$ & -0.83 $\pm.04$ & 172 $\pm15$ & B1V & 92.39 $\pm.02$ \\
(H06662) \\
EVR-HSD-009 & 4662244900292269312 & 75.0706 & -65.8073 & 15.090 $\pm.017$ & -- & -- & -- & -- & B? & 9.7143 $\pm.0003$ \\
(H17223) \\
EVR-HSD-010 & 3052089556807897600 & 107.7860 & -7.1754 & 10.222 $\pm.007$ & 14,900 $\pm100$ & 4.32 $\pm.05$ & -2.47 $\pm.14$ & 85 $\pm15$ & B5V/HB? & 110.277 $\pm.004$ \\
(H16764) \\
EVR-HSD-014 & 5393463597805768320 & 162.6241 & -39.7614 & 12.000 $\pm.002$ & 18,460 $\pm640$ & 3.21 $\pm.07$ & -1.17 $\pm.03$ & $<30$ & B3III & 34.6236 $\pm.0004$\\
(H04749) \\
EVR-HSD-015 & 5339813237197617152 & 170.0517 & -57.3402 & 12.978$\pm.001$ & 12,760 $\pm270$ & 4.25 $\pm.07$ & -2.12 $\pm.25$ & $<40$ & B8V & 20.2213 $\pm.0001$ \\
(H10605) \\
EVR-HSD-018 & 5947381578621154432 & 267.0839 & -49.6796 & 11.810 $\pm.002$ & 26,790 $\pm200$ & 3.80 $\pm.02$ & -0.78 $\pm.04$ & $<70$ & B1V & 5.39506 $\pm.00001$\\
(H05499) \\
EVR-HSD-019 & 6731538703008060288 & 285.2468 & -35.4992 & 13.142 $\pm.001$ & -- & -- & -- & --& B? & 3.41941 $\pm.00001$\\
(H16409) \\
EVR-HSD-021 & 4220013936233389184 & 301.6089 & -6.5474 & 11.409 $\pm.001$ & 28,860 $\pm100$ & 4.09 $\pm.02$ & -0.82 $\pm.08$ & 192 $\pm10$ & B0V & 2.56304 $\pm.00001$\\
(H02701) \\
EVR-HSD-024 & 6090959238631872512 & 211.1865 & -49.2117 & 11.897 $\pm.001$ & 24,270 $\pm240$ & 4.23 $\pm.03$ & -0.88 $\pm.02$ & $<60$ & BHB? & 2.27081 $\pm.00001$ \\
(H10104) \\
\multicolumn{10}{l}{\bf{Non-variables in the ES LCs}}\\
(H06720) & 5794556403014699392 & 236.5023 & -72.8776 & 13.97 & 18,170 $\pm100$ & 4.04 $\pm.02$ & -1.40 $\pm.05$ & 182 $\pm10$ & B3V & --\\
 \\
 \\
 \\
 \\
 \\
 \\
 \\
 \\
 \\
 \\
 \\
 \\
 \\
 \\
 \\
 \\
 \\
 \\
 \\
 \\
 \\
 \\
 \\
 \\
 \\
 \\
 \\
 \\
 \\
 \\
 
\end{tabular}
\label{tab:var_discoveries_sum}
\end{sidewaystable*}


\clearpage

\section{DISCUSSION} \label{section_discussion}


\subsection{Survey Sensitivity} \label{section_rates}

To test the survey sensitivity, we combine the estimated detection efficiency shown in Figure \ref{fig:det_sims_plots}, the transit fraction, and total survey targets. This offers visibility to the number of likely targets for a range of periods, and for a particular transit type (HW Vir systems, HSD / gas giant planets, and WD / planets). In all cases the survey is target limited. As demonstrated below, the survey is most sensitive to HW Vir systems (given the favorable transit likelihood and short periods) and least sensitive to long period planets.

\subsubsection{HW Vir systems} \label{section_HW_Vir_sen}

\begin{figure}[htb]
\includegraphics[width=.92\columnwidth]{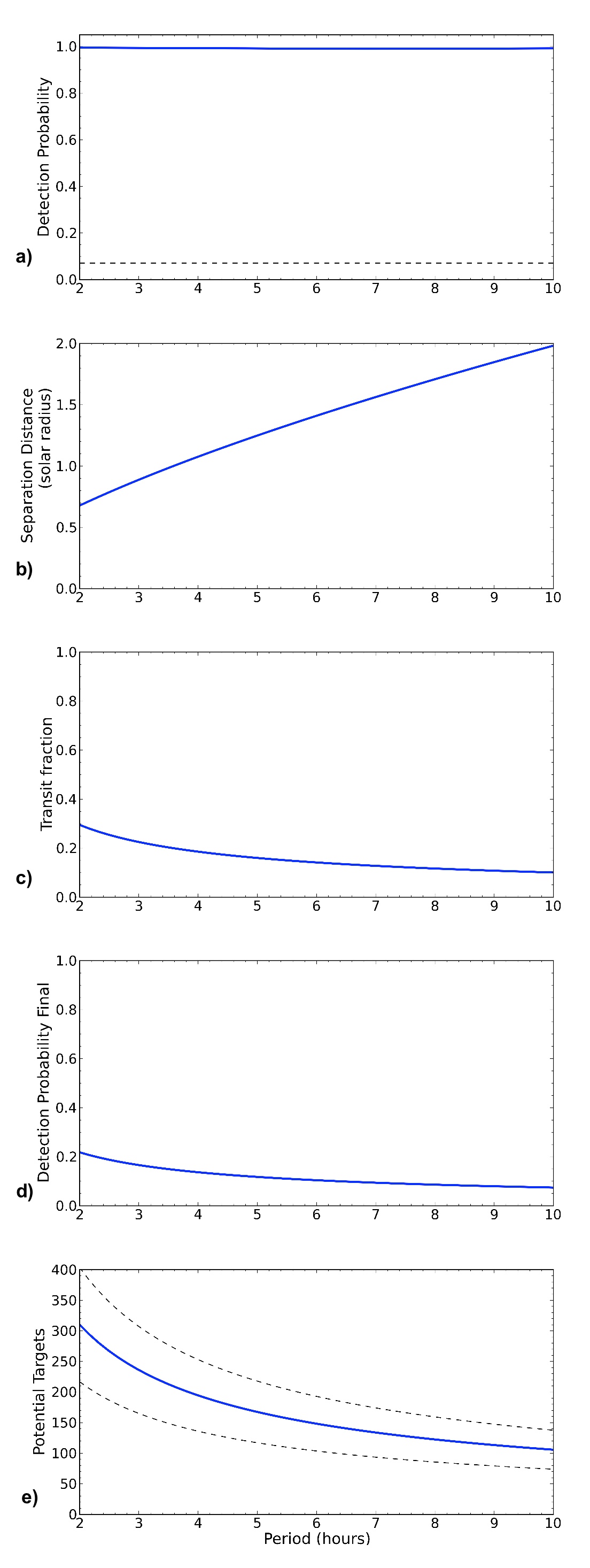}
\caption{HW Vir survey sensitivity. \textit{(a)} The detection efficiency estimated from the recovery of HW Vir like transit signals injected into Evryscope light curves (inclination angle $i=90^\circ$). The high return is the result of the fast period, many epochs, multi-year data, and high cadence light curves. The noise floor is indicated by the dashed line. \textit{(b and c)} The theoretical separation distance and transit fraction (see \S~\ref{section_HW_Vir_sen}). \textit{(d)} The final detection probability, calculated by multiplying (a) and (c) with a few adjustments for systematics (again see \S~\ref{section_HW_Vir_sen}). \textit{(e)} The potential targets that HW Vir systems could be detected in, found by multiplying (d) by the estimated total number of HSDs in the survey. The dashed lines are the estimated 1$\sigma$ errors.}
\label{fig:survey_sen_1}
\end{figure}

From the estimated detection efficiency (determined from transits simulated on to actual Evryscope light curves, see \S~\ref{section_alg_perf} and Figure \ref{fig:det_sims_plots}), we limit the period range to 2 - 10 hours, and assume a HSD primary with 0.5 $M_{\odot}$ and 0.2 $R_{\odot}$, with a companion of 0.10 $M_{\odot}$ and 0.15 $R_{\odot}$, given the parameters of the known, solved systems \citep{2018A&A...614A..77S}. The detection efficiency of the HSD survey for HW Vir systems is shown in panel (a) of Figure \ref{fig:survey_sen_1} along with the noise floor. This assumes the inclination angle $i=90^\circ$. The theoretical separation distance ($a$) and the transit fraction (using $R_{HSD}/a$) are shown in panels (b) and (c). From the detection efficiency, we subtract the noise floor and assume a $20\%$ reduction due to reduced signals from blended sources in the Evryscope pixels, difficult observing fields that affect the pipeline, or other systematics that reduce light curve quality or algorithm effectiveness. The final detection probability is shown in panel (d); and using the estimated number of total HSDs in the survey ($1422\pm{428}$ see \S~\ref{section_sum_tar}) we show in panel (e) the potential targets that we could detect HW Vir systems. We take the average over this narrow period range to be the potential targets $= 165\pm{50}$.

\subsubsection{HW Vir Occurrence Rate Estimation}

We detected seven HW Virs in our HSD survey (2 new and 5 known see Table \ref{tab:hw_virs}), including all 5 of the known systems in the declination (Dec $<+10$) and magnitude ($m_g<15$) range of the survey. Using the findings from the previous section, the frequency ($7/165$) is $4.3\%\pm{0.6}\%$ HW Vir systems in HSDs.

\clearpage

\subsubsection{HSD Planet Transits} \label{section_HSD_planet_sen}

\begin{figure}[htb]
\includegraphics[width=.90\columnwidth]{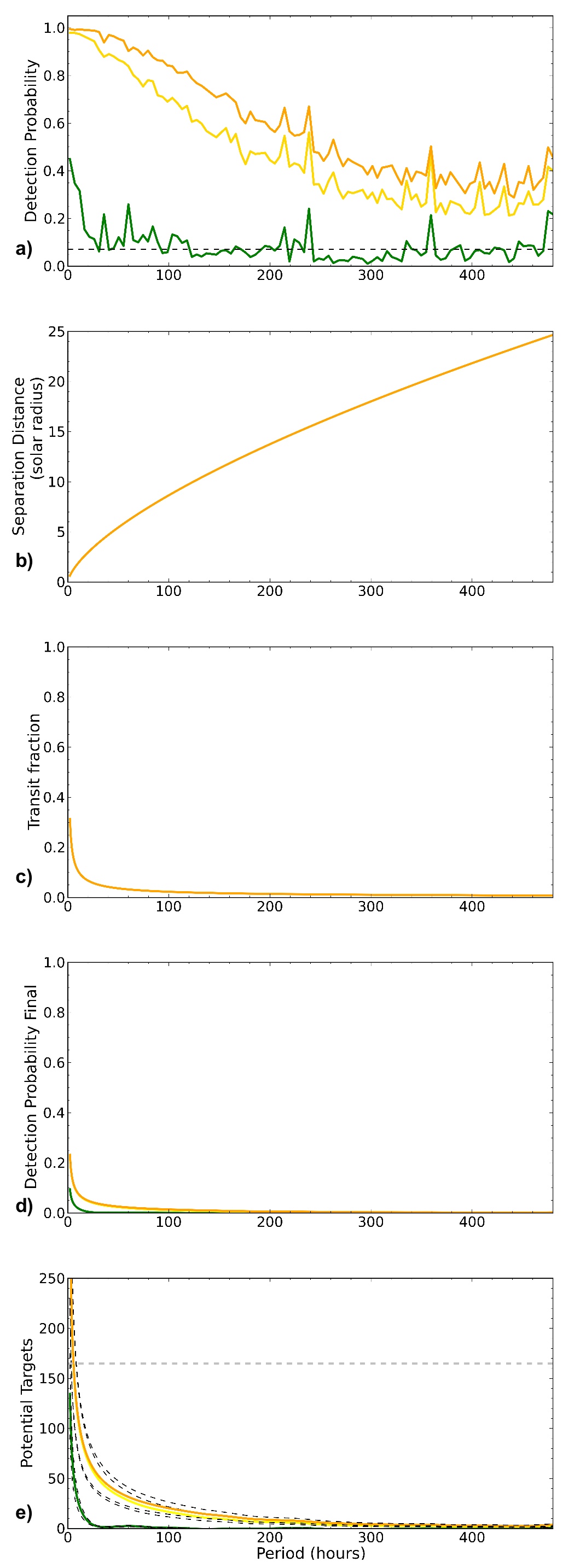}
\caption{HSD planet survey sensitivity. \textit{(a)} The detection efficiency for Super-Jupiter (orange line), Jupiter (yellow line) and Neptune (green line) planets transiting HSDs (inclination angle $i=90^\circ$). \textit{(b and c)} The theoretical separation distance and transit fraction. \textit{(d)} The final detection probability is driven down significantly for higher periods by the larger separation distance and resulting transit fraction. \textit{(e)} The potential targets that transiting planets could be detected in, found by multiplying (d) by the estimated total number of HSDs in the survey. The dashed lines are the estimated 1$\sigma$ errors. Here we keep the scaling for comparison to the different systems (HW Vir) and between different components (recovery versus transit fraction). Further in the manuscript we discuss the limiting factors for the survey and show increased detail over the period range.}
\label{fig:survey_sen_2}
\end{figure}

Using the estimated detection efficiency over the full period range of the survey (2-480 hours), we calculate the recovery rates for Super-Jupiter (5 $M_{J}$ and 0.125 $R_{\odot}$), Jupiter, and Neptune size planets transiting a canonical HSD. The gas-giant planets are recovered over the full range of the survey, while the recovery of the smaller planets decreases with increasing periods (as shown in Figure \ref{fig:survey_sen_2}). Using the same prescription as the HW Vir systems, see \S~\ref{section_HW_Vir_sen}, we calculate the separation distance, transit fraction, and final detection probability as shown in panels (b-d). The final detection probability is completely dominated by the transit fraction, and falls off significantly for periods longer than $\sim$ 20 hours. Here we keep the scaling for comparison to the different systems (HW Vir) and between different components (recovery versus transit fraction). Further in the manuscript we discuss the limiting factors for the survey. Again using the estimated number of total HSDs in the survey ($1422\pm{428}$ see \S~\ref{section_sum_tar}) we show in panel (e) the potential targets that we could detect HSD transiting planets.

\subsubsection{HSD Planet Transits Survey Sensitivity}

We detected 3 potential transiting planets, later confirmed to be other objects (see \S~\ref{section_planet_transit_candidates}). It is well known that exoplanet transit surveys suffer from high false positive rates, for example the very successful HATNet and HATSouth surveys have discovered $\sim$140 substellar objects with $\sim$ 2300 false positives as of early 2018 \citep{2018arXiv180100849B}. Although no transiting HSD planets have been discovered yet, and consequently the false positive rate is not known, there is no indication the HSD planet false positive should be particularly different than false positives for planets orbiting main sequence stars. The culprits are still likely to be eclipsing binaries or misidentified star types (HW Virs, Cataclysmic Variables, or A or B-stars in the case of HSDs). Perhaps more importantly, the instrumentation and light curve challenges that drive false positives (blended sources due to coarse pixels, crowded fields, background contamination, bad pixels, grazing eclipses, and other factors) for the Evryscope are similar to other transit surveys including the HAT instruments.

Assuming a similar false positive rate ($\sim$ 1 in 20 planet candidates will be confirmed), we would ideally want at least a few hundred potential targets that we could detect HSD transiting planets to have a decent chance of discovery. The estimated potential targets that we could detect HSD transiting gas giant planets (from \S~\ref{section_HSD_planet_sen}) are shown in Figure \ref{fig:survey_sen_3}. The potential targets are above 100 only for the very short periods for the large planets. We also show the number of targets estimated in the HW Vir analysis (the silver dashed line), providing a comparison point since we detected 7 HW Vir systems (2 new and 5 known) in the survey. Since the fall-off in potential targets is dominated by the transit fraction (see the previous section), the survey is constrained by the number of total HSD targets. In \S~\ref{section_HSD_survey_2} we show that by increasing the total survey targets by a factor of $\sim$5 would improve the potential targets that we could detect HSD transiting planets nearer to desired levels over a wider range of periods.

\clearpage

\subsection{Contribution of Blended Sources}

Additional likely HSDs targets are expected to be included in the survey as blended sources. In \S~\ref{section_blended_sources} we estimated an additional 265 blended HSD sources that can potentially contribute to the search. EVR-CB-002 offers insight as to the usefulness of these types of sources, as it is an HW Vir discovery with a nearby bright star in the field. The transit signal was reduced from 50\% to 8\% due to the blended sources (a combination of the 11.5 magnitude nearby bright star and the 13.5 magnitude target star). The transit signal from the HW Vir system and favorable inclination angle is near the deepest we would expect, and the reduced signal is near our detection limit. Given the average magnitude difference of 3 for blended Evryscope sources, EVR-CB-001 is a representative example. Thus although we did discover this system, it seems likely that we would not recover signals with a more grazing eclipse or from smaller transiting objects. For HW Vir systems if we assume an $\approx$25\% recovery, this still only gives an additional 65 targets - minor compared to the total targets and well below the estimated error range.

\subsection{Compact Binaries} \label{section_compact_binary_discussion}

In this section, we consider compact binaries showing a light curve variation due to ellipsoidal deformation, with an asymmetric shape due to gravitational limb darkening and Doppler beaming. The unseen companion is assumed to be a WD, but could potentially be a more compact object, while the primary is a HSD or HSD like in color-magnitude space and spectral features. In the HSD survey, we found 3 of these systems (1 known and 2 discoveries) out of 1422 likely HSD targets. The recoveries were found with BLS, with relatively low power near the survey average, and one of the detections was at a half-period alias. LS missed two and found one at half the period, the Outlier detector is not designed for these signals and did not recover any of the systems. The systems we recovered all showed amplitudes above 5\%, and at least a 1\% difference in even versus odd depths.

The detection of this type of system faces the difficulties of the HSD search (fast timescale variability and a limited number of targets), but with the added challenge of discriminating the asymmetric shape from the common sinusoidal like variable. The typical failures are for the matched filter to either miss the variability or find the half-period alias, or for the reviewer to not recognize the multi-component variability and mistake the object for an unexceptional variable. 

The HSD survey in this work was designed to search for a variety of variable signals over a wide period range. A subsequent search concentrating on very short periods only (10 minutes to 10 hours), with more aggressive systematics removal for all variability longer than 10 hours, and with a custom detection algorithm designed specifically for the unique asymmetric even / odd cycle light curve could potentially recover additional systems. We leave that search for future work.

We can only make a rough estimate of the occurrence rate given the subjectivity in the recovery ability, and low number of discoveries. As the limiting factor for detection in this work is the size of the difference in even versus odd depths, we require this to be 1\% or more which means the main amplitude in the light curve variation to be $\approx$ 5\% or more. Even with an inclination of 63 degrees, EVR-CB-001 (see \S~\ref{section_discovery}) has an amplitude well above this. We assume these systems are detectable in Evryscope light curves up to a 45 degree inclination, and that our detection efficiency is less than the HW Vir systems but still reasonably high at $\approx$ 0.8 (given the very short periods and many periods captured in the light curves). The estimated frequency is: $3 / (45/90 \times 0.8 \times 1422) = 0.005$. Less than a half percent of HSDs are likely to be compact binary systems with a HSD like primary and unseen companion.

\subsubsection{HSD Survey 2} \label{section_HSD_survey_2}

\begin{figure}[htb]
\includegraphics[width=1.0\columnwidth]{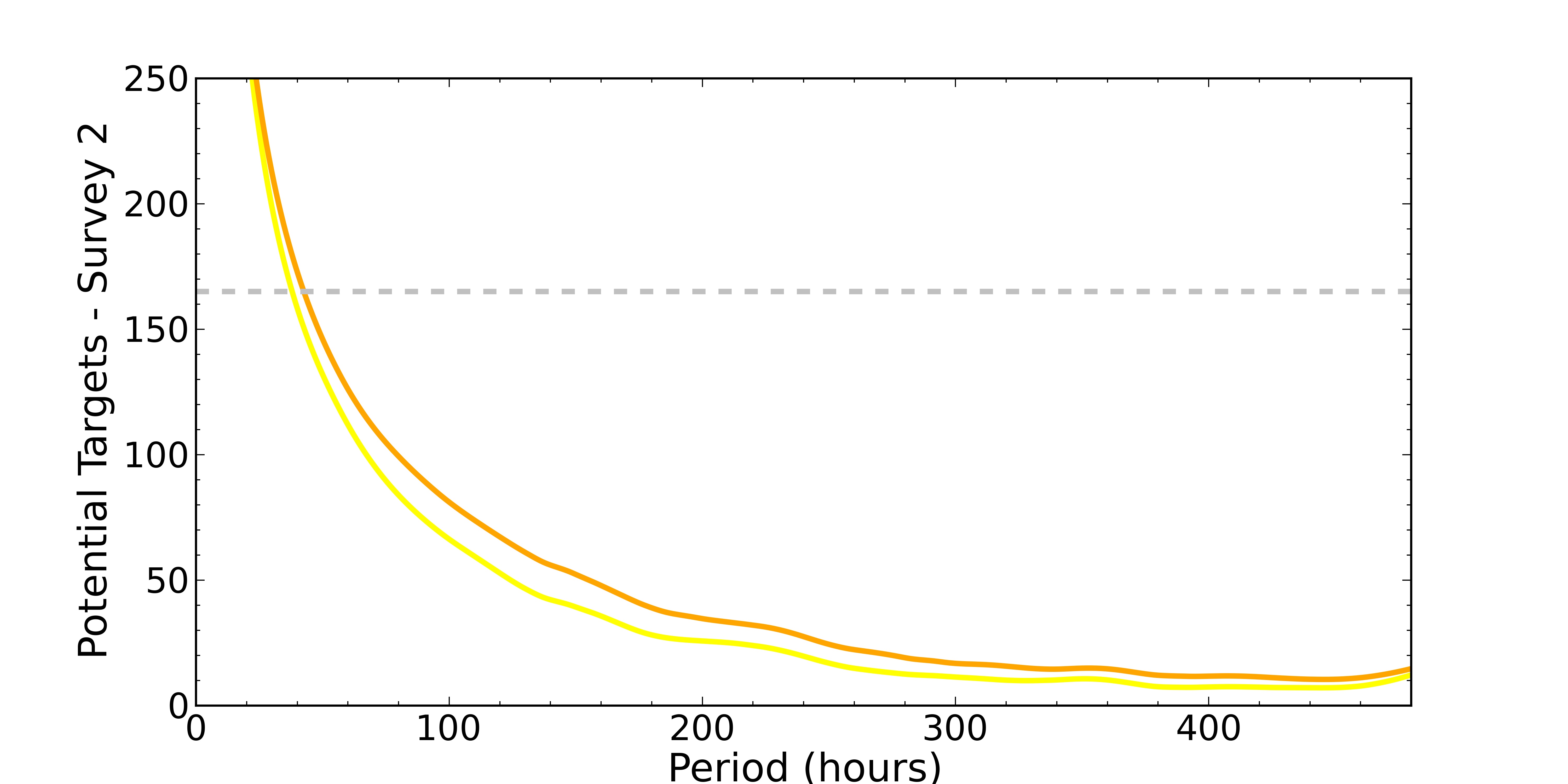}
\caption{The potential targets that transiting planets could be detected in, Super-Jupiter size (orange line) and Jupiter size planets (yellow line) are shown. The Survey 2 with increased magnitude and FOV coverage increases the potential detectable transit targets to nearly 100 for periods up to 100 hours.}
\label{fig:survey_sen_3}
\end{figure}

The survey in this work is comprised of southern sky targets with magnitudes brighter than 15.0M in $m_{g}$. Based on the Geier based GAIA HSD list, including stars to $m_{g} < 16$ approximately doubles the number of targets. The Evryscope North (a copy of the CTIO system) was deployed to Mount Laguna Observatory (MLO) in late 2018. In two years time it will have collected a similar number of epochs for a similar number of sources as the CTIO data the survey in this work was based off. This would approximately double again the number of HSD targets. In Figure \ref{fig:survey_sen_3}, we show the effect of the increased survey scope. The potential targets with detectable planets is at or above 100 out to periods of $\approx$ 100 hours for super-Jupiter and Jupiter size planets, and is favorable for Neptune size planets to at least several days. We would also expect on order tens of targets over the full test period range.

The increased scope survey (Evryscope HSD Survey 2, Ratzloff, et al., in prep), is expected to find a similar fraction of rare, fast transit HSD systems including compact binaries, HW Virs, and reflection binaries, but at an increased total yield of $\approx$ 4 times driven by the increased number of targets. From Figure \ref{fig:survey_sen_2}, we demonstrate the combination of Evryscope 2 minute cadence, photometry, and detection algorithms are effective at recovering potential HSD transiting planet signals. The recovery of actual planet candidates in this work, even though they are false positives, further validates the survey performance. Relative to the planet search, with the increased targets we should be well placed to explore the more untapped regions past the very short periods with sensitivity to potentially make discoveries. 

\section{SUMMARY} \label{section_summary}

We conducted an all southern sky survey of bright HSDs searching for fast transit signals in the Evryscope light curves. The Evryscope data is 2 minute cadence, with continuous all southern sky observing for multiple years. We estimate the number of HSD targets in this work to be approximately 1400. Based on our recovery rates from transit simulations and the fraction of transiting objects, we expected to be sensitive to HSD variability of different types including compact binaries, HW Vir systems, transiting planets, reflection binaries, and other variables. We discovered 14 new HSD variables including 2 very rare compact binaries with unseen WD companions, 2 bright HW Virs, several reflection effect binaries and sinusoidal variables. Four of the systems are published in separate discovery papers solving the system parameters in detail. We also discovered 24 other variables in the survey including several post GB, HB, and BHB variable systems. We obtained spectra for the discoveries and determined the spectral types, and we identified the discoveries that are good candidates for future followup. A planned followup survey expanding the targets in this work by at least a factor of 4 is discussed.


\section*{Acknowledgements}

This research was supported by the NSF CAREER grant AST-1555175 and the Research Corporation Scialog grants 23782 and 23822. HC is supported by the NSFGRF grant DGE-1144081. BB is supported by the NSF grant AST-1812874. The Evryscope was constructed under NSF/ATI grant AST-1407589. This work was supported by the National Science Foundation through grant PHY 17-148958.

This work includes observations obtained at the Southern Astrophysical Research (SOAR) telescope, which is a joint project of the Minist\'{e}rio da Ci\^{e}ncia, Tecnologia, Inova\c{c}\~{o}es e Comunica\c{c}\~{o}es (MCTIC) do Brasil, the U.S. National Optical Astronomy Observatory (NOAO), the University of North Carolina at Chapel Hill (UNC), and Michigan State University (MSU).

This research has used the services of \mbox{\url{www.Astroserver.org}} under reference EVRY01. P.N. acknowledges support from the Grant Agency of the Czech Republic (GA\v{C}R 18-20083S).

This paper includes data collected by the TESS mission. Funding for the TESS mission is provided by the NASA Explorer Program.

This research has made use of the SIMBAD database, operated at CDS, Strasbourg, France.

This research has made use of the VizieR catalogue access tool, CDS, Strasbourg, France.

This research has made use of the International Variable Star Index (VSX) database, operated at AAVSO, Cambridge, Massachusetts, USA.


\bibliographystyle{apj}
\bibliography{ft_survey}


\clearpage
\appendix

\begin{table}[h]
\caption{Variables\\(Likely A or B-stars misclassified as HSDs)}
\centering
\begin{tabular}{ l c c c c c}
ID & ID & RA & Dec & mag [G] & Period [h]\\
\hline
New Discoveries\\
EVRJ044348.48-854516.6 & GaiaDR24614260804078737792 & 70.9520 & -85.7546 & 10.587 $\pm.001$ & 35.9327 $\pm.0004$\\
EVRJ052817.76-690418.5$^a$ & GaiaDR24658105788836255360 & 82.0740 & -69.0718 & 11.272 $\pm.002$ & 474.7 $\pm.6$ \\
EVRJ053824.72-663523.6$^b$ & GaiaDR24659534879024982528 & 84.6030 & -66.5899 & 13.941 $\pm.002$ & 36.287 $\pm.003$ \\
EVRJ065540.80-234417.5 & GaiaDR22922396976293672576 & 103.9200 & -23.7382 & 10.396 $\pm.001$ & 21.6214 $\pm.0001$\\
EVRJ071431.63-60949.0 & GaiaDR23058298056594337920 & 108.6318 & -6.1636 & 12.753 $\pm.001$ & 55.335 $\pm.001$ \\
EVRJ072950.66-133935.3$^c$ & GaiaDR23033287603040592896 & 112.4611 & -13.6598 & 12.691 $\pm.001$ & 166.199 $\pm.009$ \\
EVRJ074738.21+053614.4 & GaiaDR23137857549744826752 & 116.9092 & 5.6040 & 14.167 $\pm.001$ & 3.60397 $\pm.00001$ \\
EVRJ075018.70-252635.5 & GaiaDR25602331396479925120 & 117.5779 & -25.4432 & 13.829 $\pm.001$ & 509.3 $\pm.8$ \\
EVRJ075521.12-163622.3 & GaiaDR25718296303734937856 & 118.8380 & -16.6062 & 13.618 $\pm.001$ & 57.997 $\pm.001$ \\
EVRJ080955.39-461701.3 & GaiaDR25519602045642615808 & 122.4808 & -46.2837 & 11.029 $\pm.001$ & 19.9735 $\pm.0001$ \\
EVRJ081242.38-180840.6 & GaiaDR25719913582258286592 & 123.1766 & -18.1446 & 13.243 $\pm.001$ & 196.78 $\pm.01$ \\
EVRJ090801.42-461506.1 & GaiaDR25327604500572716928 & 137.0059 & -46.2517 & 10.476 $\pm.001$ & 112.444 $\pm.004$ \\
EVRJ150433.67-170155.2 & GaiaDR26305932286056884480 & 226.1403 & -17.0320 & 13.584 $\pm.001$ & 10.90413 $\pm.00003$ \\
EVRJ155252.37-645012.5$^d$ & GaiaDR25825969999996438656 & 238.2182 & -64.8368 & 12.814 $\pm.006$ & 21.1933 $\pm.0001$ \\
\hline
\multicolumn{6}{l}{$^a$(noted as a potential super giant in \cite{2012ApJ...749..177N}, and as an unidentifiable variable in \cite{2002AcA....52..397P}),}\\
\multicolumn{6}{l}{$^b$(noted as a potential OB star in \cite{1970CoTol..89.....S}),}\\
\multicolumn{6}{l}{$^c$(noted as a potential OB star in \cite{1971PWSO...1a...1S}),}\\
\multicolumn{6}{l}{$^d$(noted as a unidentifiable variable ASASSN-V J155246.87-644843.7 in \cite{2018MNRAS.477.3145J})}\\

\end{tabular}
\label{tab:var_discoveries_3}
\end{table}

\subsection{Discovery Comments}

\subsubsection{EVRJ150433.67-170155.2, EVRJ155252.37-645012.5}

Periodic transit like features are visible in these variables with no ellipsoidal effects or secondary eclipses evident. The transit durations are too long (2-3 hours) for the primary to be a HSD, but too short to be an O or B star. The 10.9041 and 21.1933 hour period variables could be CVs or Novae; they require further followup to reveal the characteristics of the systems.

\subsubsection{EVRJ072950.66-133935.3}

A 166.1992 hour long period, very eccentric EB with $\approx$ 9 hour eclipse durations. Both primary and secondary are reasonably deep (0.26 primary). We suspect the system is likely comprised of O and B stars, making this a potentially rare eclipsing binary with very hot and massive components. Again the bright magnitude will aid in followup.

\clearpage

\begin{figure*}[htb]
\includegraphics[width=0.95\columnwidth]{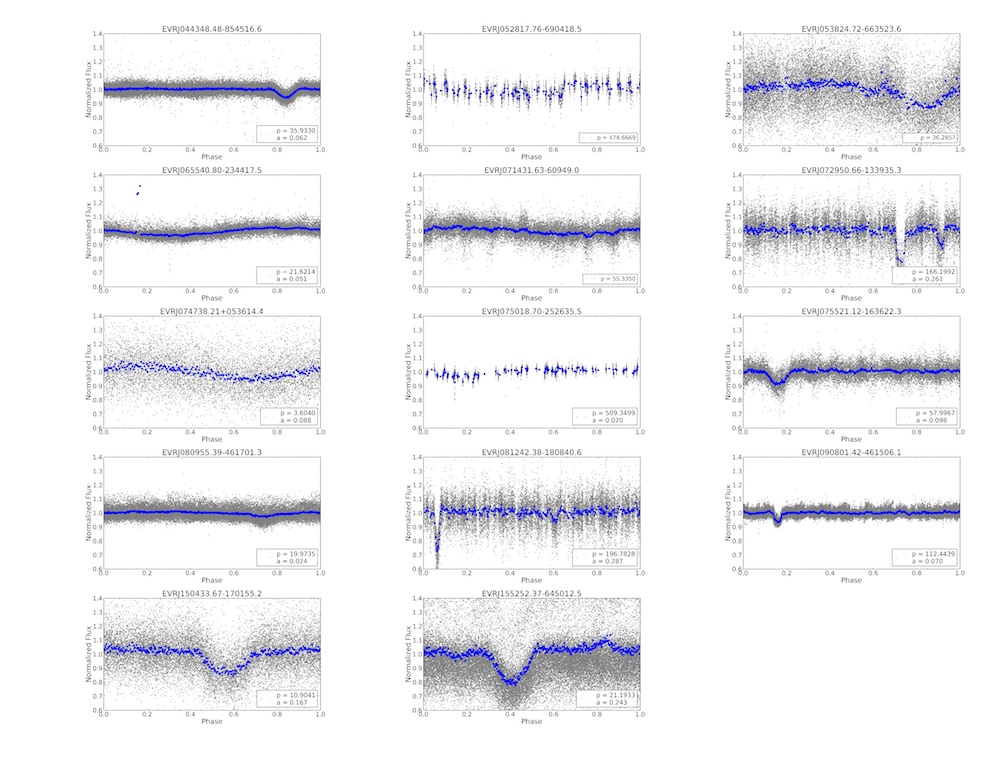}
\caption{The Evryscope light curves of variable discoveries (likely A stars or other stellar types) showing variable signals with periods ranging from a few hours to several months. The period and amplitudes shown are from the best LS fit for the sinusoidal variables, and for the eclipsing binaries a Gaussian is fit to the primary eclipse to measure the depth. Grey points = 2 minute cadence, blue points = binned in phase.}
\label{fig:var_disc_other_1}
\end{figure*}

\end{document}